\documentclass{aa}  
\usepackage{linenoaa}
\usepackage{graphicx}
\usepackage{txfonts}

\usepackage{booktabs}
\usepackage{titlesec}
\usepackage[dvipsnames]{xcolor}
\usepackage{subcaption}
\usepackage{pdflscape}
\usepackage{placeins}

\usepackage{color}
\usepackage{natbib,twoopt}
\usepackage[hyphenbreaks]{breakurl}
\usepackage[breaklinks]{hyperref}      
\bibpunct{(}{)}{;}{a}{}{,}          
\definecolor{cobalt}{rgb}{0.06, 0.2, 0.65}
\hypersetup{
  colorlinks,
  citecolor=blue,
  linkcolor=blue,
  urlcolor=blue,
}
\makeatletter
  \newcommandtwoopt{\citeads}[3][][]{\href{http://adsabs.harvard.edu/abs/#3}%
    {\def\hyper@linkstart##1##2{}%
     \let\hyper@linkend\@empty\citealp[#1][#2]{#3}}}
  \newcommandtwoopt{\citepads}[3][][]{\href{http://adsabs.harvard.edu/abs/#3}%
    {\def\hyper@linkstart##1##2{}%
     \let\hyper@linkend\@empty\citep[#1][#2]{#3}}}
  \newcommandtwoopt{\citetads}[3][][]{\href{http://adsabs.harvard.edu/abs/#3}%
    {\def\hyper@linkstart##1##2{}%
     \let\hyper@linkend\@empty\citet[#1][#2]{#3}}}
  \newcommandtwoopt{\citeyearads}[3][][]%
    {\href{http://adsabs.harvard.edu/abs/#3}
    {\def\hyper@linkstart##1##2{}%
     \let\hyper@linkend\@empty\citeyear[#1][#2]{#3}}}
\makeatother

\titleformat{\paragraph}
{\normalfont\normalsize}{\theparagraph}{1em}{}
\titlespacing*{\paragraph}
{0pt}{3.25ex plus 1ex minus .2ex}{1.5ex plus .2ex}

\newcommand{\bse}{BSE\xspace}
\newcommand{\mocca}{MOCCA\xspace}
\newcommand{\fewbody}{FEWBODY\xspace}

\titlerunning{Sky Localization of DWDs inside GCs}
\authorrunning{Y. Zhao et al.}

\begin{document} 

   \title{Distance Estimation and Sky Localization of Eccentric Double White Dwarf Binaries from Gravitational Wave Observations inside Globular Clusters}

   \author{
    Y. Zhao$^{1}$, L. Hellström$^{2}$, S. Zhang$^{3}$,
    M. Giersz$^{2}$, A. Askar$^{2}$, Y. Lu$^{4,5}$, Z. Cao$^{1}$, D. Belloni$^{6}$, G. Wiktorowicz$^{2}$, M. Górski$^{2}$
    }

   \institute{School of physics and astronomy, Beijing Normal University, 19 Xinjiekouwai St, Beijing 100875, People’s Republic of China
   \email{ytzhao@bnu.edu.cn}
   \and
   Nicolaus Copernicus Astronomical Center, Polish Academy of Sciences, ul. Bartycka 18, PL-00-716 Warsaw, Poland
   \and
   Department of Physics, Centre for Extragalactic Astronomy, Durham University, South Road, Durham DH1 3LE, UK
    \and 
    School of Astronomy and Space Sciences, University of Chinese Academy of Sciences, 19A Yuquan Road, Beijing 100049, People’s Republic of China
    \and
    National Astronomical Observatories, Chinese Academy of Sciences, 20A Datun Road, Beijing 100101, People’s Republic of China
    \and
    International Centre of Supernovae (ICESUN), Yunnan Key Laboratory of Supernova Research, Yunnan Observatories, CAS, Kunming 650216, China
    }

   \date{Accepted XXX. Received YYY; in original form ZZZ}

  \abstract{The cosmic distance scale is built on multiple different techniques for estimating distances in space that are often connected and dependent on multiple measurements and assumptions. Double white dwarf binaries (DWDs) are common objects and are expected to produce gravitational wave (GW) signals that can be observed with space-based detectors such as LISA. By analyzing these signals we should be able to estimate the distance and sky location of the source. Previous studies have done this for circular binaries which, while they are abundant, have, in general, weaker signals than eccentric binaries and it is not possible to differentiate whether a circular binary is in the field or in a dense environment such as a globular cluster (GC). In this paper we used eccentric binaries from MOCCA GC simulations, simulated the GW signal from each binary at locations related to GCs in the Milky Way and estimated the precision on the distance and the sky location of the source. We find that distances can be estimated with higher precision than current day methods even with low eccentricity binaries and higher eccentricity further increases this precision. Although the probability of finding a tight and eccentric DWD is far lower than a circular one, we can expect to find at least a few in the dense environments of the Milky Way, such as GCs. These estimations would be independent measurements with high precision to objects inside dense environments, such as GCs inside the Milky Way and the Magellanic Clouds.} 


   \maketitle

%

\section{Introduction}

White dwarfs (WDs) are very abundant and are often found in binaries. With the introduction of space-based telescopes, such as LISA \citep{LISA_2017, lisa_2024}, Taiji \citep{Ruan2018}, and TianQin \citep{TianQin_2016, TianQin_2025}, they are a prime candidate for low-frequency gravitational wave (GW) observations. The large number of double white dwarfs (DWDs) suggests that their weak signals may collectively contribute significantly, or even dominate, the gravitational wave foreground \citep{nelemans2001,ruiter2010,Robson2017,Thorpe2019,korol2022,Amaro2023, Zhang2025}. DWDs are also expected to be observable beyond the Milky Way, with LISA capable of resolving systems throughout the Local Group and satellite galaxies \citep{2018ApJ...866L..20K,2020ApJ...894L..15R,2020A&A...638A.153K}. Unresolved populations may contribute to an extragalactic stochastic background \citep{2024MNRAS.531.2642R}, while precision measurements of their signals could probe Galactic dynamics \citep{2020ApJ...901....4B,2025PhRvD.111d4023E}.

White dwarfs and DWDs are commonly found in dense environments such as globular clusters (GCs), where dynamical interactions between single stars and binaries can affect the binary properties, dissolve the binary, or form new binaries. This may create binaries with features unique to these environments, which are not found in the field \citep{hellstrom2025} where stars and binaries evolve in isolation. GCs are particularly interesting since they are much denser than open clusters (OCs), leading to more dynamical interactions, and studies have suggested that in certain GCs, there is a WD subsystem in the center of the cluster \citep{kremer2021, strickler2009, calamida2008} where the probability for a DWD to be involved in a dynamical interaction is high. In addition to this, GCs are very old, which means that even stars less massive than the Sun have time to evolve into WDs compared to a younger cluster where only heavier stars have had time to evolve into WDs. 

Distances to globular clusters can be determined using a variety of techniques that differ both in their physical basis and in their achievable precision. Broadly, those methods are divided into two categories: geometric or semi-geometric methods and standard candles.

Geometric and semi-geometric methods include trigonometric parallaxes, eclipsing binaries, Baade-Wesselink analyses applied to RR Lyrae or Type II Cepheids and kinematic methods based on comparing line-of-sight velocity dispersions with proper-motion measurements. Trigonometric parallax is very efficient at measuring distances to close-by stars. However, the precision of this method is dependent on the distance to the source and thus it is not efficient at larger distances. Furthermore, this method requires the source to be optically distinguishable and thus cannot be used in dense regions, such as, on the other side of the galactic bulge in the Milky Way (MW). Parallaxes can reach distance precisions of $1-5$\% using Gaia EDR3 \citep{Gaia_D3, maiz2021, gaia2023} and $<2$\% using the Very Long Baseline Interferometry \citep{maiz2021}. However, the latter requires bright radio sources, limiting the number of potential sources. 

Eclipsing binaries can be used to measure distances by observing the light curve and radial velocities of the stars. The precision of this method is good, on the order of $<5$\% \citep{helminiak2013} down to less than $2$\% \citep{pietrzynski2013}. The practical limitation of this approach is that we need extensive, high-quality photometric (optical and near-infrared) and spectroscopic observations for suitable late-type systems, which are rare and difficult to identify in globular clusters. The ultimate precision is limited by the intrinsic scatter of the color-surface brightness relation. For the blue stars the situation is worse due to lower precision of radial-velocity measurements and larger scatter of the surface brightness- (blue) color relation.

The Baade-Wesselink method uses the rapid pulsations of stars to determine the distances to them. Recent analysis of RR Lyrae and Type II Cepheids have shown this to be a promising method of obtaining distances to GCs \citep{zgirski2025}. Kinematic methods use line-of-sight velocity dispersions combined with proper-motion measurements \citep{baumgardt2021}.

Standard candles are used to measure distances by relying on objects with known brightness. The precision of these distance estimates are good and reliable at large distances and are often used to measure cosmic distances. In practice, young clusters at larger distances, such as in the Large Magellanic Cloud (LMC) and Small Magellanic Cloud (SMC) could host Cepheid variables or binaries involved in supernova\,Ia. Limitations in this method include reliance on other distance measurements for calibration, dust and extinction will affect the results and need to be properly accounted for. Most important standard candles are period-luminosity relation for RR Lyrae stars and the brightness of tip of the red giant branch (TRGB). Other methods, such as horizontal-branch fitting, color-magnitude diagram modeling or Type II Cepheids, generally do not provide a substantial improvement in precision or are limited in practice by their low numbers in globular clusters.

In practice, a key goal is to calibrate selected distance indicators in globular clusters using distances obtained from independent geometric or semi-geometric methods, such as parallaxes or kinematics. This is particularly important for establishing an accurate calibration of the TRGB in the optical (I) band and in the near-infrared, because the TRGB can be used instead of Classical Cepheids to measure distances to galaxies hosting Type Ia Supernovae and form complementary/alternative Cepheid based distance ladder:

\begin{itemize}
    \item Eclipsing binaries (Gaia calibration in MW) $\rightarrow$ Cepheids (Large Magellanic Cloud distance with
eclipsing binaries) $\rightarrow$ SN Ia host galaxies (Cepheid distances).
    \item TRGB in Globular clusters $\rightarrow$ SN Ia host galaxies (TRGB distances).
\end{itemize}

However, to calibrate the TRGB we need a few clusters with different metallicity to calibrate the color brightness relation for near-infrared bands, and at least 100 stars in the bright part of the RGB ( approximately 1 mag below TRGB, because of statistical precision), for each cluster or a composite CMD for a few clusters. Currently, the distance to GC is not precise enough.

The electromagnetic (EM) emission from DWDs can be detected by optical surveys such as \textit{Gaia} and Vera C. Rubin Observatory \citep{Gaia2016, LSST}. Previous optical studies have shown that eclipsing binary systems are relatively easy to identify, and have yielded of order one thousand detached DWD candidates in the Galaxy \citep{Jin2025, Korol2019}. Accreting WDs, such as AM\,CVn systems and cataclysmic variables are also expected to be numerous, especially in dense star clusters. To date, two candidate AM CVn-type binaries and $\sim10^2$ of cataclysmic variables have been identified electromagnetically in GCs \citep{Hurley2002, Andronov2003, Pala2020, Sandoval2018, Zurek2016, Green2025}. However, observable EM counterparts are limited to a small nearby subset of systems, primarily because most DWDs have intrinsically weak EM emission and current surveys lack the depth and cadence to detect faint or infrequent accretion signatures; dust extinction in the Galactic disk further suppresses detectability along some lines of sight. In contrast, GWs offer an independent probe that is unaffected by the environment and can reveal systems invisible to EM surveys. A multi-messenger approach combining GW and EM observations is therefore highly valuable. It can break degeneracies, enable secure source identification, and yield stronger constraints on population properties and binary evolution.

Despite the large number of circular and tight DWDs in relation to eccentric and tight ones \citep{vaNZeist2025_b}, eccentric binaries are more useful in our study. The strength of the GW signal is largely dependent on eccentricity and we will show in this paper that even a small increase in eccentricity will drastically increase the precision of distance and localization estimations from GWs. In addition, since tight and eccentric DWDs are unlikely to form in the field, we can, with quite good confidence, claim that such a binary is inside a dense environment if observed. Because of that we focus this study and paper on tight and eccentric DWDs inside GCs.

Current methods of measuring astronomical distances are limited and most depend on additional measurements, information about the sources, or assumptions. By observing GW signals from eccentric DWDs we could have independent and accurate estimates for both distance and localization inside the MW which could offer an additional and fundamentally different way to test and refine the cosmic distance scale. The aim of this paper is to investigate whether using the GW signals from DWDs inside dense environments will give accurate distance and localization estimates. It is purely a computational paper, aimed at exploring this idea, not proposing objects for observations. The paper is structured as follows. In section \ref{sec:formChannels} we present formation channels of tight and eccentric DWDs and in section \ref{sec:modelsAndInitialConditions} we describe the models and initial conditions used. Section Section \ref{sec:GWObs} describes the pipeline used to obtain the GW signal and to estimate parameter precision. Section \ref{sec:results} presents the results of these simulations, and Section \ref{sec:discussion} discusses their implications, summarizes our conclusions, and outlines directions for future work.



\section{Formation channels of tight and eccentric DWDs}
\label{sec:formChannels}
Using GW signals to distinguish the origin of a DWD (field or dense environment) is very challenging. Most observable parameters are not unique to either origin and cannot be used to determine its background. Tight DWDs in the field are almost exclusively circular since the orbital separations of their progenitors are small enough so that tidal forces circularize the binary before or during the common-envelope phase \citep{nandez2015, wilson2020}. However, tight DWDs in dense environments can be eccentric due to dynamical interactions \citep{hellstrom2025}. It is theoretically possible for a triple system to form a tight and eccentric DWD as the inner binary \citep{rajamuthukumar2025}. However, as argued by \cite{rajamuthukumar2025} the triple channel is expected to be very unlikely and for most triple systems, the tertiary star is too distant to affect the inner binary in a significant way. This results in eccentricity being an indication of a DWD from a dense environment. Eccentric binaries are preferable due to the GW signal being stronger and easier to differentiate above the GW background signal. However, the number of tight and eccentric DWDs is far fewer than tight and circular.

There are a few different channels to produce a tight and eccentric binary in a dense environment. For primordial binaries, the DWD needs to form with a small semi-major axis, and during the common-envelope phases, the binary circularizes, resulting in a tight and circular binary, regardless of initial eccentricity. After the formation of the DWD, dynamical interactions may increase the eccentricity of the binary. This dependence on dynamical interactions after the DWD is formed is the reason why the field is unlikely to host any tight and eccentric binaries. It is possible that the eccentricity of a tight DWD is increased inside a dense environment and then the binary is kicked out into the field. However, due to the very small lifetimes of these tight and eccentric DWD, this is very unlikely. In our data this does not happen for any model.

For dynamically formed binaries we have two different channels; the binary is formed directly with two WDs and high eccentricity or the binary is formed with any type of stars, later form a DWD due to stellar evolution and is then involved in additional dynamical interactions that may increase the eccentricity. These two channels can happen as a result of binary formation or an exchange in a dynamical interaction. For more details on the formation channels of both primordial and dynamically formed binaries see \cite{hellstrom2025}. 


\section{Models and initial conditions}
\label{sec:modelsAndInitialConditions}

\subsection{The \mocca code}
\label{sec:moccaCode}

In order to get a realistic population of binaries in GCs we use the MOnte Carlo Cluster simulAtor (\mocca\footnote{\url{http://moccacode.net}}) code \citep{MOCCA1,MOCCA2, hypki2025, giersz2025} to evolve GCs and extract the binaries from the models in the MOCCA-SURVEY Database III \citep{hypki2025, giersz2025}. \mocca is built on a Monte Carlo approach and is much faster than N-body codes while still retaining good accuracy for the evolution of GCs over a Hubble time. Previous studies have shown that these models are good approximations for MW GCs \citep[e.g.][]{hypki2025, giersz2025} and the properties of the DWDs inside them are in good agreement with the observed \citep{hellstrom2024}. Dynamical interactions are solved with the \fewbody code \citep{freqeau2004, freqeau2007} for both binary-binary and binary-single interactions.

The prescriptions for stellar evolution is important for the WD population since it may affect the properties of stars and binaries. \mocca uses a modified version of the \bse code \citep{Hurley2000, Hurley2002} which originally consists of a set of algorithms that prescribes the evolution of both single and binary stars. The modifications to the code \citep[see, e.g.,][for more details]{kamlah2022, giersz2025_b} that affects this work is improved wind prescriptions \citep{belczynski2016}, treatment of the common-envelope phase \citep{giacobbo2018}, proper prescription for cataclysmic variable evolution \citep{belloni2018b, belloni2019} and remnant mass prescriptions for black holes (BHs) and neutron stars \citep{banerjee2020}. 

\subsection{Initial conditions}
\label{sec:initialConditions}
\subsubsection{GC models}
\label{sec:models}
The models that we use are the same as in \cite{hellstrom2024, hellstrom2025}, that is, 185 GC models that survive longer than 9 Gyr. For a full description of the initial conditions see \citet{hypki2025}. Compared to the traditional setup of GC models with one stellar populations, these models contain multiple stellar populations (MSP) that are introduced at $t=0$ without any time delay. The two different populations are vastly different. The first population is tidally filling and has a higher maximum initial stellar mass. The second population is tidally underfilling, more compact, and has a lower maximum initial stellar mass. Due to the properties of the first population, it loses a large amount of mass (approximately $30-40$\%) during the first few Myr and thus has a large number of escapers. 

The most important parameters for the GC models are the following:
\begin{itemize}
    \item A range of initial cluster masses, from $5.5\times10^{5}$ to $2.2\times10^{6}$ stellar objects, modeled with two stellar populations. Note that binaries are counted as one object. 
    \item Both tidally filling and tidally underfilling initial cluster configurations.
    \item Two initial binary fractions: 10\% and 95\%.
    \item A Kroupa mass function between 0.08 and 150 M$_{\odot}$ (population 1) and 0.08 and 20 M$_{\odot}$ (population 2).
    \item Different Galactocentric distances between 1 and 4 kpc.
    \item Different properties of MSPs (for details see \citealt{hypki2025}).
\end{itemize}

Previous studies have found that core-collapsed clusters are supported by WD subsystems and DWDs \citep{kremer2021}. Due to the reliance on dynamical interactions to form these binaries (see Sect.~\ref{sec:formChannels}), the core of the clusters is a probable location for the formation of such binaries. However, the presence of an intermediate-mass BH (IMBH) or a BH subsystem will lower the population of DWDs in the dense regions and thus fewer DWDs will be involved in dynamical interactions. We investigate the models that produced at least one eccentric DWD at ${t>9}$\,Gyr and find that out of 56 models we have a BH subsystem in 22 models, an IMBH in 33 models and 1 model contains neither at $12$\,Gyr. This model, lacking both a BH subsystem and an IMBH, has undergone core collapse and is supported by a WD subsystem. However, it only produced one tight and eccentric DWD during its lifetime, showing no increase in efficiency for producing this type of binary. This may be caused by the increased number of dynamical interactions disrupting primordial binaries that would otherwise evolve into tight binaries.

For all our models (185 in total) we find that 127 contain a BH subsystem, 43 contain an IMBH and 15 contain neither and are core collapsed at $12$\,Gyr. This means that we find at least one eccentric DWD in 56 out of 127 models that contain a BH subsystem, in 33 out of 43 models that contain an IMBH, while only 1 out of 15 core-collapsed models managed to produce such a binary. 

\subsubsection{Binaries}
\label{sec:binaries}
We extract a few binaries with different eccentricities from the models described in Sect.~\ref{sec:models} and their parameters can be found in Table~\ref{tab:binaries}. These binaries are representative of the larger population and were picked based on their eccentricity, semi-major axis and mass. We selected one circular binary where both members are heavy ($>1$\,M$_{\odot}$) and the semi-major axis is $\sim0.5$\,R$_{\odot}$.

In addition to this we select 3 eccentric binaries with low ($\sim$0.18), high ($\sim$0.82) and very high ($\sim$0.97) eccentricity. All of these eccentric binaries are tight and produce relatively strong GW signals compared to most DWDs. The circular binary was placed in each cluster from the \citet{2018MNRAS.478.1520B} catalog, see Table~\ref{tab:gcs2} for the parameters, the low and high $e$ binaries were placed at the 50 heaviest GCs within $10$\,kpc of the Sun, see the bolded rows in Table~\ref{tab:gcs2} and the very high $e$ binary was simulated in 10 GCs within $10$\,kpc; 4 clusters close to the Sun, 3 not very close clusters and 3 clusters close to 8\,kpc. These are marked with a star in Table\,\ref{tab:gcs2}. The very high eccentricity binary was simulated in fewer clusters due to limits on computation time.

\begin{table}
    \caption{Table of the 4 binaries used in our simulations.}              
    \label{tab:binaries}     
    \centering                               
    \begin{tabular}{c c c c c}
    \hline\hline
        Label & $M_1$ [M$_{\odot}$] & $M_2$ [M$_{\odot}$] & a [R$_{\odot}$] & e \\
        \hline
            Circular binary & 1.342 & 1.121 & 0.508 & 0.0   \\
            Low $e$           & 0.725 & 0.227 & 0.247 & 0.178 \\
            High $e$          & 1.238 & 1.00  & 0.671 & 0.822 \\
            Very high $e$     & 0.713 & 0.55  & 1.022 & 0.967 \\
        \hline
    \end{tabular}
\end{table}

These binaries were selected to span the characteristic parameter space in eccentricity and separation relevant for detectable sources, rather than to represent a statistically complete population sample. They should therefore be interpreted as illustrative test cases for the parameter-estimation analysis.

\subsubsection{Localization and distance to real globular clusters}
In order to measure the precision of our estimations we need the location and distance to the source. The \cite{2018MNRAS.478.1520B} catalog includes the distance to the cluster ($D$), the right ascension (RA), declination (Dec), mass of the cluster and the tidal radius ($r_t$). To obtain the area in the sky for each cluster we used the tidal radius to first get the size of the cluster in the sky in degrees as
\begin{equation}
    \theta = r_t / D \cdot (180/\pi).
\end{equation}
This angle was then used to calculate the area of each cluster in the sky, i.e., 
%
\begin{equation}
    A = \pi \theta^2.
\end{equation}

\begin{figure*}
    \centering

    \begin{subfigure}[t]{0.48\textwidth}
        \centering
        \includegraphics[width=\linewidth]{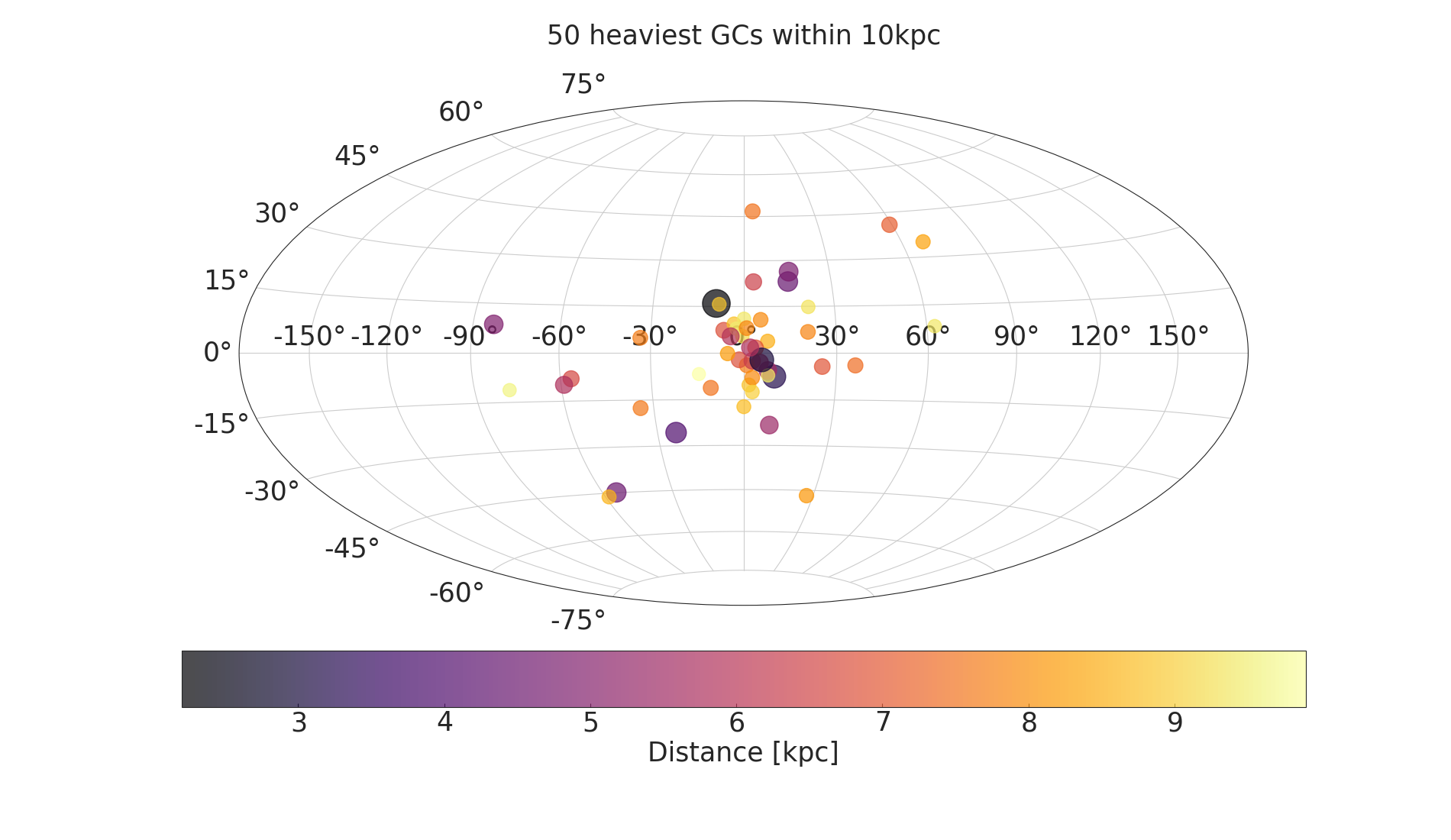}
        \label{fig:skymap_dist}
    \end{subfigure}
    \hfill
    \begin{subfigure}[t]{0.48\textwidth}
        \centering
        \includegraphics[width=\linewidth]{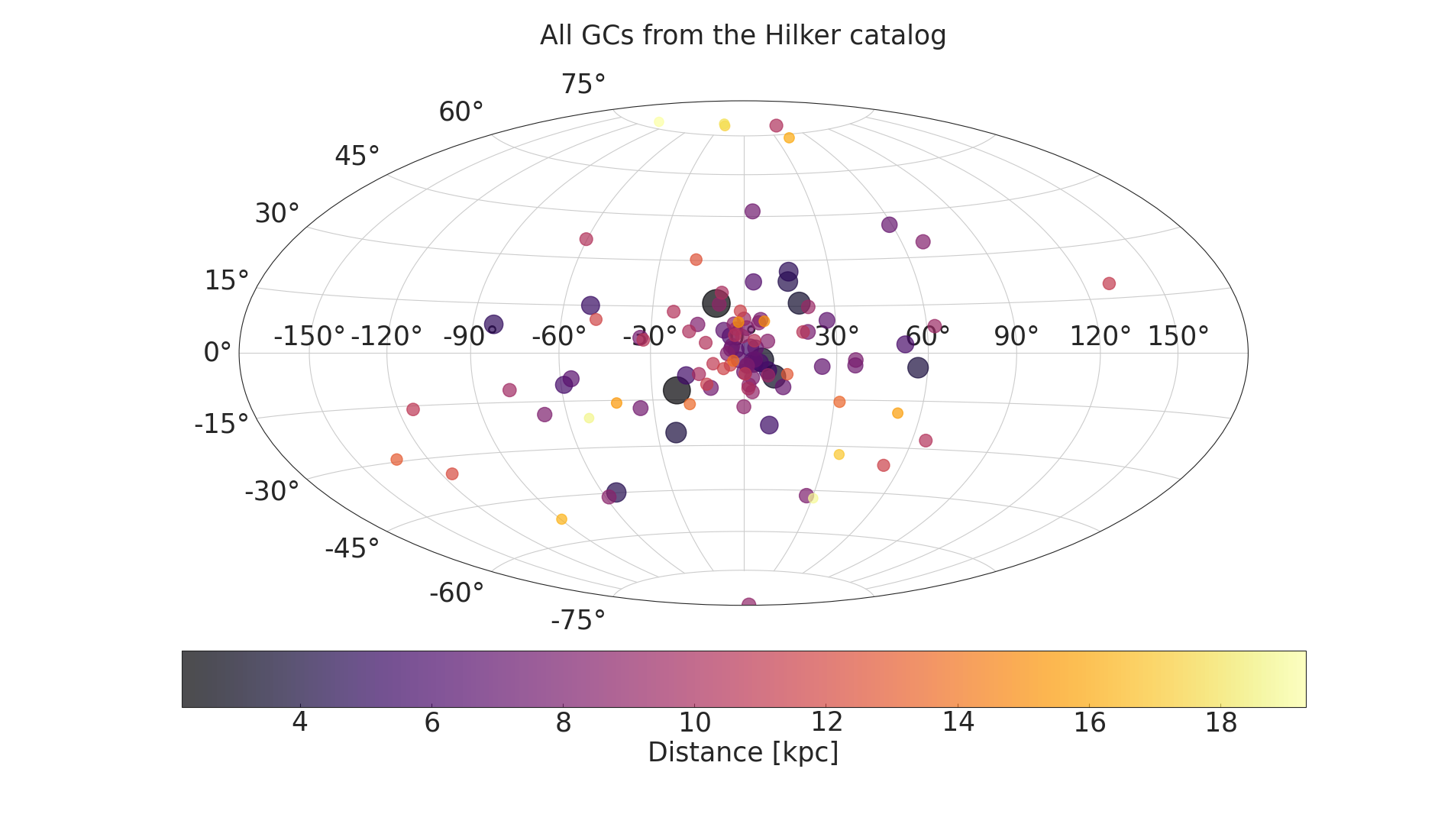}
        \label{fig:skymap_circ_dist}
    \end{subfigure}

    \caption{Skymap of the GCs location and distance to the GCs that we used for the eccentric binaries (first panel) and the circular binaries (second panel). The plot is made in galactic coordinates and the color map shows the distance to the cluster. The marker size is also related to the distance where a larger marker shows a closer cluster to the Sun.}
    \label{fig:skymapDistances}
\end{figure*}

Figure \ref{fig:skymapDistances} shows two skymaps. Panel a) shows the heavy clusters that were used for the eccentric binaries and panel b) shows all GCs in the \cite{2018MNRAS.478.1520B} catalog that were used for the circular binaries. These plots show that we have a bias towards the galactic bulge with a few outliers spread out in the sky. 

\section{GW simulations}
\label{sec:GWObs}

In this section, we describe how to calculate the GW Signal-to-Noise Ratio for DWDs and how to estimate the parameter uncertainties for those detectable DWDs.

\subsection{Signal-to-Noise Ratio estimation for eccentric DWDs}

Circular-orbit gravitational wave sources will radiate GWs at frequencies twice their orbital frequencies. Unlike the circular case, eccentric DWDs produce GW radiation at multiple frequency harmonics. The angle-averaged signal-to-noise ratio ($\varrho$) of eccentric sources detected by LISA can be estimated via the sum of contributions from all the harmonics \citep{2007ApJ...665L..59W}, that is,
\begin{equation}
	\varrho^2 = \sum_{n} \int\left[\frac{h_{c,n}(f_n)}{h_{\rm rms}(f_n)}\right]^2 d{\rm ln}f_n,
	\label{eq:rho}
\end{equation}
where $f_n = nf_{\rm orb}$ is the $n$-th harmonic GW frequency, $f_{\rm orb}$ is the orbital frequency of DWDs, $h^2_{\rm rms}(f_n) = (20/3)S_h(f_n)f_n$ is the sky averaged root-mean square noise of GW detectors, $S_h(f_n) $ is noise spectrum of LISA \citep{2019CQGra..36j5011R}, $h_{c,n}$ is the characteristic strain of the $n$-th harmonic for eccentric DWDs with the following form
\begin{equation}
	h_{c, n} = \frac{1}{\pi D}\sqrt{\frac{2G}{c^3}\frac{dE_n}{df_n}},
\end{equation} 
where $dE_n/df_n$ is the GW energy emitted at different harmonics which is given in the work of \citet{1964Peters}. Considering the frequency band swept by each DWD source, we multiply $h_{c,n}$ by $N = {\rm min}[1, \dot{f_n}T_{\rm obs}/f_n]$ in Eq.~\eqref{eq:rho}. For circular binaries, the GW radiation contains only the $n=2$ harmonic, and the above formula simplifies considerably, as shown by \citet{2015CQGra..32e5004M}.

\begin{figure}
    \centering
    \includegraphics[width=0.8\linewidth]{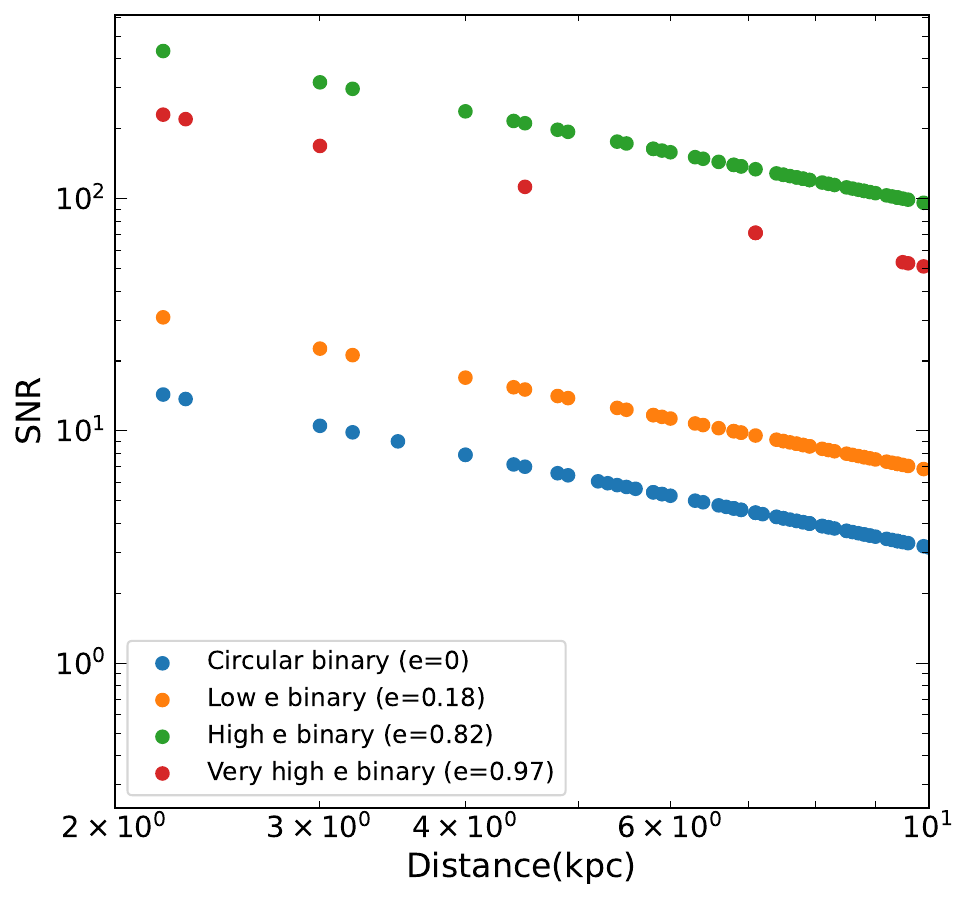}
    \caption{Scatter plot of SNRs and distances for 4 different DWDs observed by LISA over 4 years and located in different GCs. Blue, orange, green, and red colors represent the circular binary ($e = 0$), low $e$ binary ($e = 0.18$), high $e$ binary ($e = 0.82$), and very high $e$ binary ($e = 0.97$), respectively.}
    \label{fig:snr}
\end{figure}

For the $4$ selected binaries listed in Table~\ref{tab:binaries}, we calculated the signal-to-noise ratio (SNR) values detected by LISA for their locations in various real clusters over a $4$-year observation period. Fig.~\ref{fig:snr} illustrates the SNR values for $4$ DWDs located at different clusters within $10$\,kpc distance apart from us. DWDs located at farther clusters have smaller SNR. Circular DWD shows the smallest SNR value among the four sources. This difference comes mainly from the peak radiation frequency of circular binary and eccentric binaries. Circular DWD located at the nearest cluster has SNR higher than $10$, and for this particular binary it can be successfully detected by LISA when located within $4$\,kpc. The variation in SNR among the three eccentric DWDs arises from differences in both eccentricity and binary separation. We find that the two DWDs with eccentricities larger than $0.8$ could be detected by LISA out to $10$\,kpc with high SNR values. For highly eccentric DWDs located at nearby clusters, the SNR values could be larger than $200$. For the low eccentricity DWD sample, it can be detected by LISA with SNR\,$\geq8$ out to a distance of ${\sim8.3}$\,kpc.

\subsection{Sky localization and distance estimation of DWDs}
\label{sec:skyLocaAndDistResults}

Following the above SNR analyses, in order to estimate the parameter precision of DWDs, including localization and distance measurement, we adopt the commonly used Fisher Information Matrix (FIM) method \citep{zhaolu2023,zhaoyt2025} as 
\begin{equation}
	\Gamma_{ab} = 2\sum_{\alpha}\int_{0}^{T}\frac{\partial
    \hat{h}_{\alpha}(t)}{\partial\lambda^{a}}\frac{\partial \hat{h}_{\alpha}(t)}{\partial\lambda^{b}}dt,
\end{equation}
where $\hat{h}_{\alpha}$ is the noise weighted non-sky averaged waveform in $\alpha$-th independent detectors of LISA, $T=4$ years is the observation period of LISA. $\hat{h}_{\alpha}$ is calculated as
\begin{equation}
	\hat{h}_{\alpha}(t) = \sum_{n=1}^{n_{\rm max}}h_{\alpha,n}(t)/S_h^{1/2}(f_n(t)).
\end{equation}
$h_{\alpha,n}$ is the $n$-th harmonic timedomain waveform
\begin{equation}
	h_{\alpha,n}(t) = \frac{\sqrt{3}}{2D}[F^{+}_{\alpha}(t)A^{+}_{n}(t)+F^{\times}_{\alpha}A^{\times}_{n}(t)],
	\label{eq:ht}
\end{equation}
where $F^{+}_{\alpha}$ and $F^{\times}_{\alpha}$ are the plus and cross part of the antenna pattern function. $A_n^{+}$ and $A_n^{\times}$ are the $n$-th harmonic coefficients with the following explicit form:
\begin{equation}
	A_n^{+} = -(1+cos^2\iota)[a_ncos(2\gamma)-b_nsin(2\gamma)]+[1-cos^2\iota]c_n,
\end{equation}
\begin{equation}
	A_n^{\times} = 2cos^2\iota[b_ncos(2\gamma)+a_nsin(2\gamma)],
\end{equation}
where $\iota$ is the inclination angle, $\gamma$ is the angle of periastron, and the detailed form of $a_n$, $b_n$, and $c_n$ can be found in the work of \citet{2004PhRvD..69h2005B}. We also consider the doppler effect due to LISA's orbital motion. We adopt $n_{\rm max} = 20(1+e)^{1.1954}/(1-e^2)^{3/2}$ since when $n > n_{\rm max}$ the contribution of the harmonics to the GW strain amplitude can be safely neglected \citep{2003ApJ...598..419W}. From the FIM, we can get the estimated errors from the covarience matrix by:
\begin{equation}
	\Delta\lambda^{a} = \sqrt{(\Gamma^{-1})^{aa}}.
\end{equation}
We consider $6$ parameters ($\lambda = [\mathcal{M}_c, {D}, e_0, \theta_S, \phi_S, f_0]$) in our FIM analyses for eccentric binaries, which are chirp mass $\mathcal{M}_c$ of DWDs, luminosity distance $D$, eccentricity at the moment of beginning of observation by LISA $e_0$, the polar angle $\theta_S$ and azimuthal angle $\phi_S$ in the ecliptic coordinate frame converted from the RA and DEC of cluster information in Table~\ref{tab:gcs2}. For circular DWDs, since their waveform is much simpler than that of eccentric binaries, we calculate their FIM in the same way as the work of \citet{2020PhRvD.102f3021H}. For the circular case, we adopted $7$ parameters in their FIM, which are the GW strain amplitude $\mathcal{A}$, frequency at the beginning of observation by LISA $f_0$, $\dot{f}$, inclination angle $\iota$, polarization angle $\psi$, the polar angle $\theta_S$ and azimuthal angle $\phi_S$. Because of the degeneracy of chirp mass and luminosity distance, we get the estimation errors on chirp mass and distance from the formula of error propagation.

\section{Results from precision estimations}\label{sec:results}

\begin{figure}[t]
    \centering
    \begin{subfigure}[t]{\textwidth}
        \includegraphics[width=0.5\linewidth]{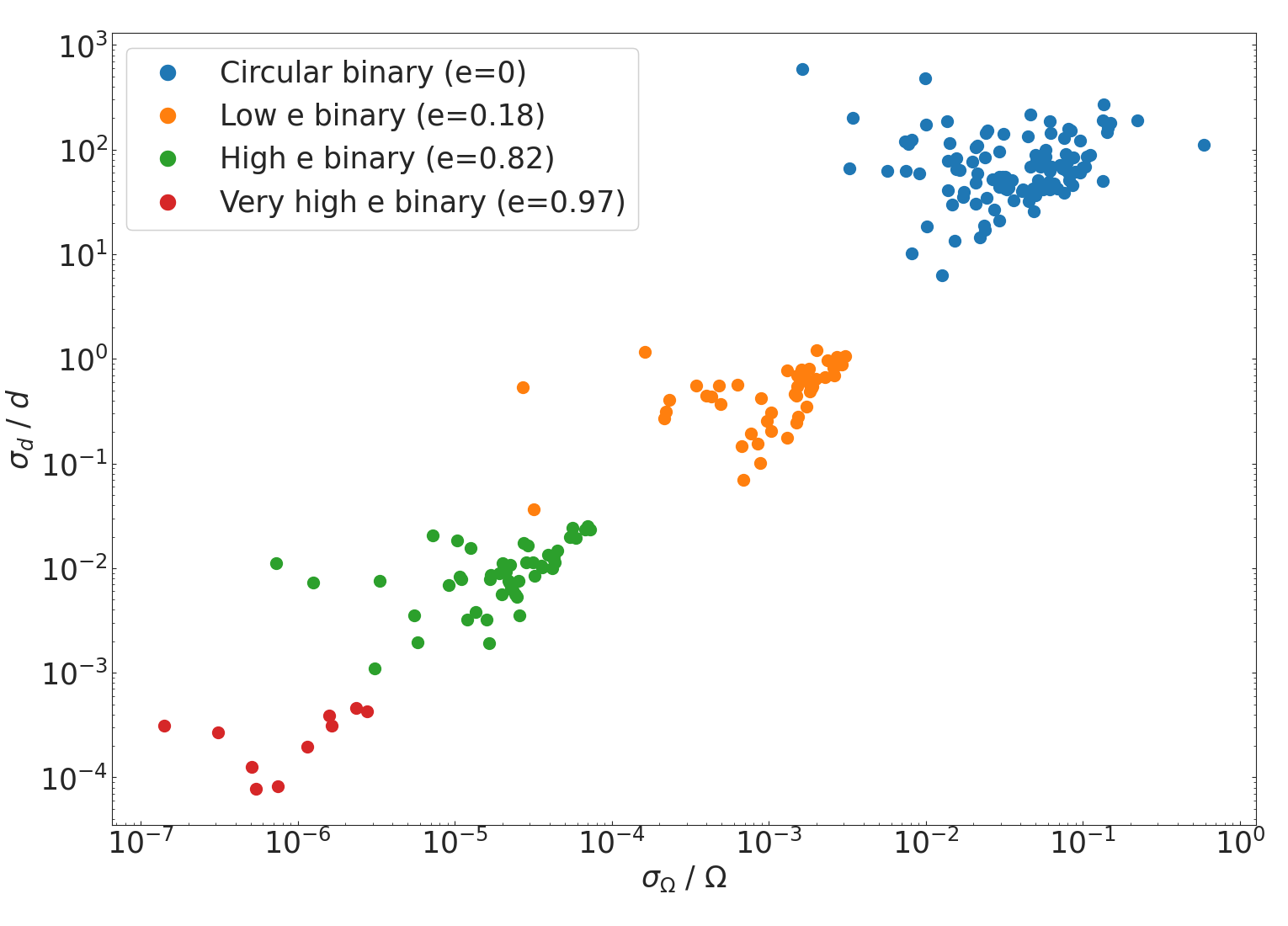}
    \end{subfigure}
    \begin{subfigure}[t]{0.5\textwidth}
        \includegraphics[width=\linewidth]{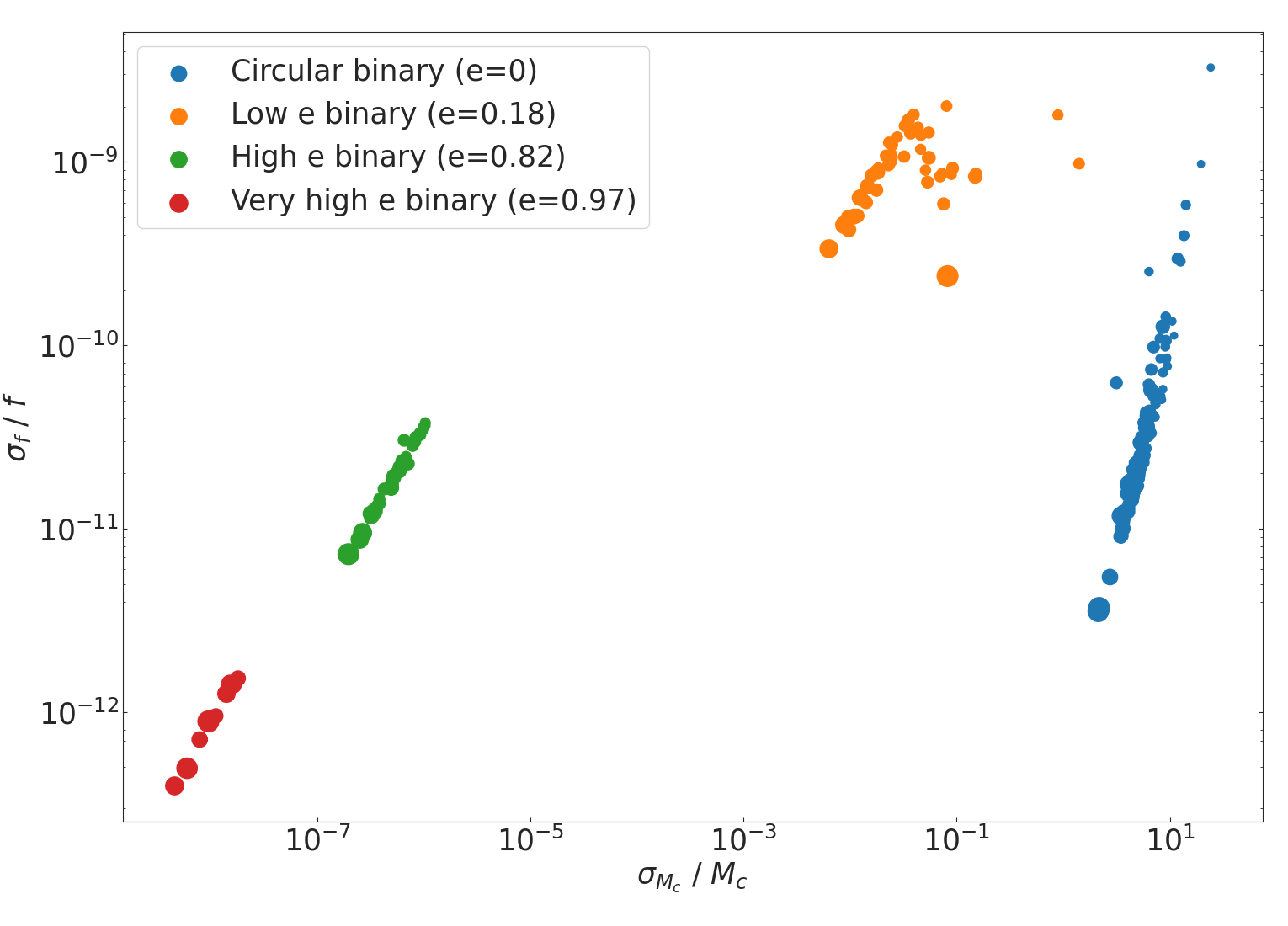}
    \end{subfigure}
    \begin{subfigure}[t]{0.5\textwidth}
        \includegraphics[width=\linewidth]{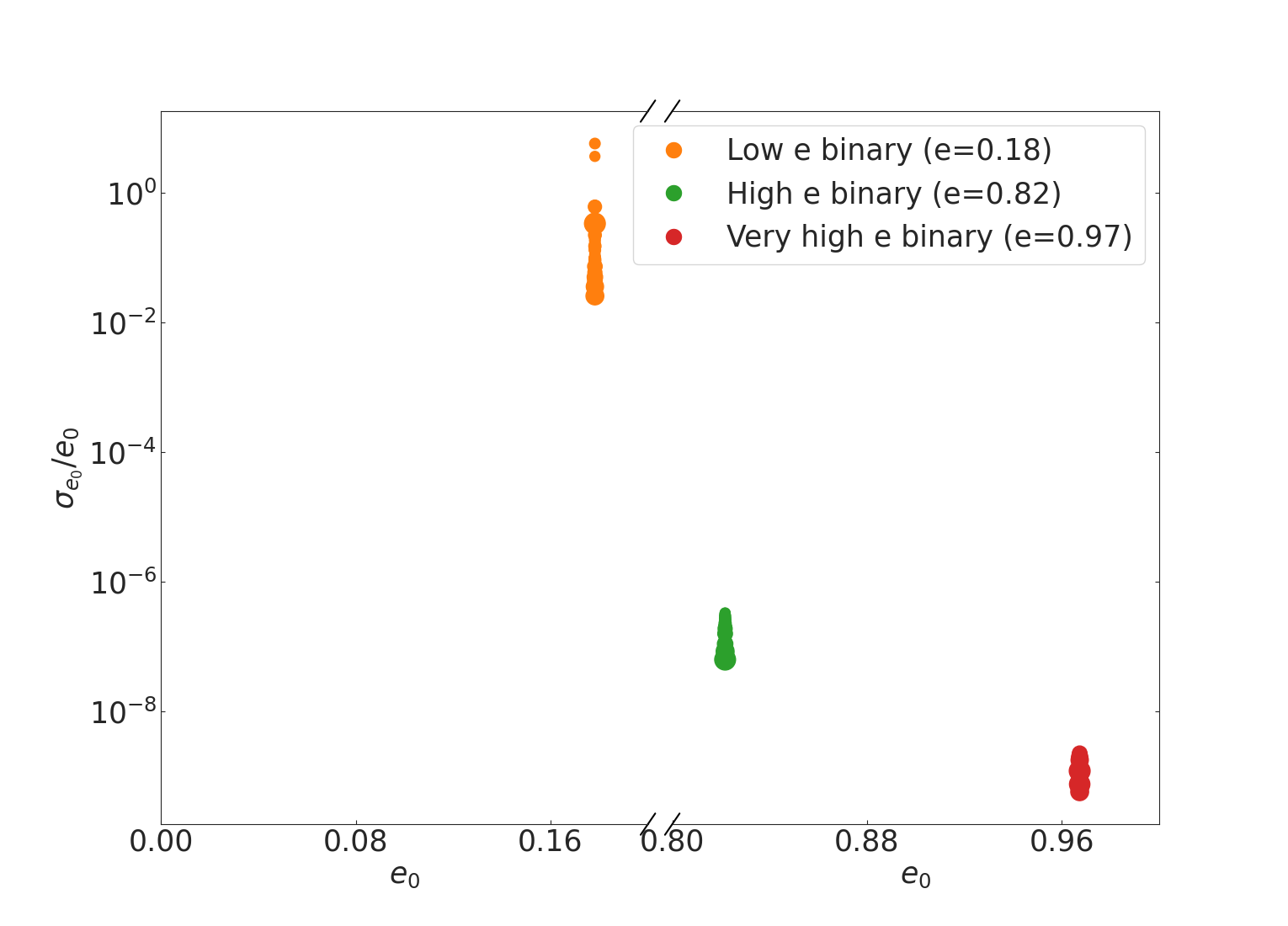}
    \end{subfigure}
    
    \caption{Scatter plots of relative errors for different variables for the 4 different binaries at each cluster location. The first panel shows the relative errors on the localization estimate plotted against that on the distance estimate. The second panel shows the relative errors on the chirp mass estimate plotted against that on the frequency estimate. The third panel shows the relative errors on the eccentricity estimate plotted against the initial eccentricity of the binary. Blue is the circular binary, orange the low $e$ binary, green the high $e$ binary and red the very high $e$ binary. The marker size shows the distance to the cluster where a larger marker means a smaller distance to the cluster.}
    \label{fig:scatterAcc}
\end{figure}

In Fig.~\ref{fig:scatterAcc}, we plot the precisions of the parameter estimations. The color corresponds to our 4 different binaries; blue for the circular binary, orange for low $e$ ($e = 0.18$) binary, green for the high $e$ ($e = 0.8$) binary and red for the very high $e$ ($e = 0.97$) binary. We can see that for all panels the groups form 4 distinct islands and are very easy to differentiate from each other. More detailed explanations are as follows.

The first panel shows the error in localization plotted against the error in distance. There are 4 distinct groups connected to each binary where the circular binary has the highest error and it is easy to see that a higher eccentricity gives a lower error on both localization and distance estimation. The precision of distance estimation for low eccentricity binary ($e = 0.18$) is between $0.02$ and $0.14$, for high eccentricity binary ($e = 0.8$) is around $0.0005$ and $0.002$, and for very high eccentricity binary ($e = 0.97$) is around $0.00003$ and $0.0001$. Compared to the eccentric binaries, circular binaries in clusters can be hardly measured with high precision, with $\sigma_{d}/d$ between $5$ and $30$. The reason of this comes from the high SNRs for eccentric binaries as well as the break of degeneracy of parameter estimations after considering eccentricity.
We show the localization for DWDs in clusters in terms of the relative sky areas which is defined as the fraction of the localization area to its host cluster area in the sky, i.e. $\sigma_{\Omega}/\Omega$.
The localization error spread for both eccentric and circular binaries exceeds that of luminosity distance, likely due to the size of the clusters. Some clusters are very small in area on the sky, both due to the size of the cluster and the distance to it, and thus even if the error is fairly small, the error as a percentage can be large. For most low eccentricity binaries, the relative error on sky areas is between $0.0001$ and $0.1$, except for one outlier, which comes from the closer cluster.
For high eccentricity binaries their $\sigma_{\Omega}/\Omega$ is between $3\times10^{-7}$ and $0.001$, and for very high eccentricity binary the relative sky localization areas are about $3\times10^{-8}-3\times10^{-5}$.

The second panel shows the relative error on chirp mass plotted against the relative error in frequency. The four binaries form 4 distinct groups and they are easy to distinguish from each other. The high and very high eccentricity binaries have the lowest errors on both parameters, compared to the low eccentricity binary. Binaries with higher eccentricity spread their GW radiation to much more harmonics of frequencies, leading to more information of the waveform, thus resulting more accurate estimations on both parameters. The much more scattered distribution of low $e$ binary in the plane of relative errors for chirp mass and frequency comes from the estimation precision on eccentricity which can be also seen from the third panel. Less eccentric sources will show stronger degeneracy in the parameter estimation process. For circular binary illustrated here, we find that their relative errors on frequency is comparable to the highly eccentric binaries, reaching to around $10^{-11}$. Since the estimation of frequency depends on the sampling rate during data process stage. For circular binaries, the relative error of chirp mass estimation is higher than the order of $1$, since the degeneracy of chirp mass and luminosity distance. While for binaries with very high eccentricity, the relative error on chirp mass could be under $10^{-7}$.

The third panel show the relative error of the initial eccentricity plotted against the initial eccentricity, which means the eccentricity value at the beginning observation of LISA,. It is easy to see that the error becomes smaller as the eccentricity gets larger. For low eccentricity binary, the relative precision of eccentricity is between $\sim0.01$ to $\sim6$, while for high eccentricity binary and very high eccentricity binary, the precision for uncertainty of eccentricity is much higher reaching around $10^{-7}$ and $10^{-9}$. The precious result of estimation on the eccentricity of DWDs by LISA shows that we could measure the highly eccentric binaries with great precision.

\begin{figure*}
    \centering
    \begin{subfigure}[t]{0.48\textwidth}
        \centering
        \includegraphics[width=\linewidth]{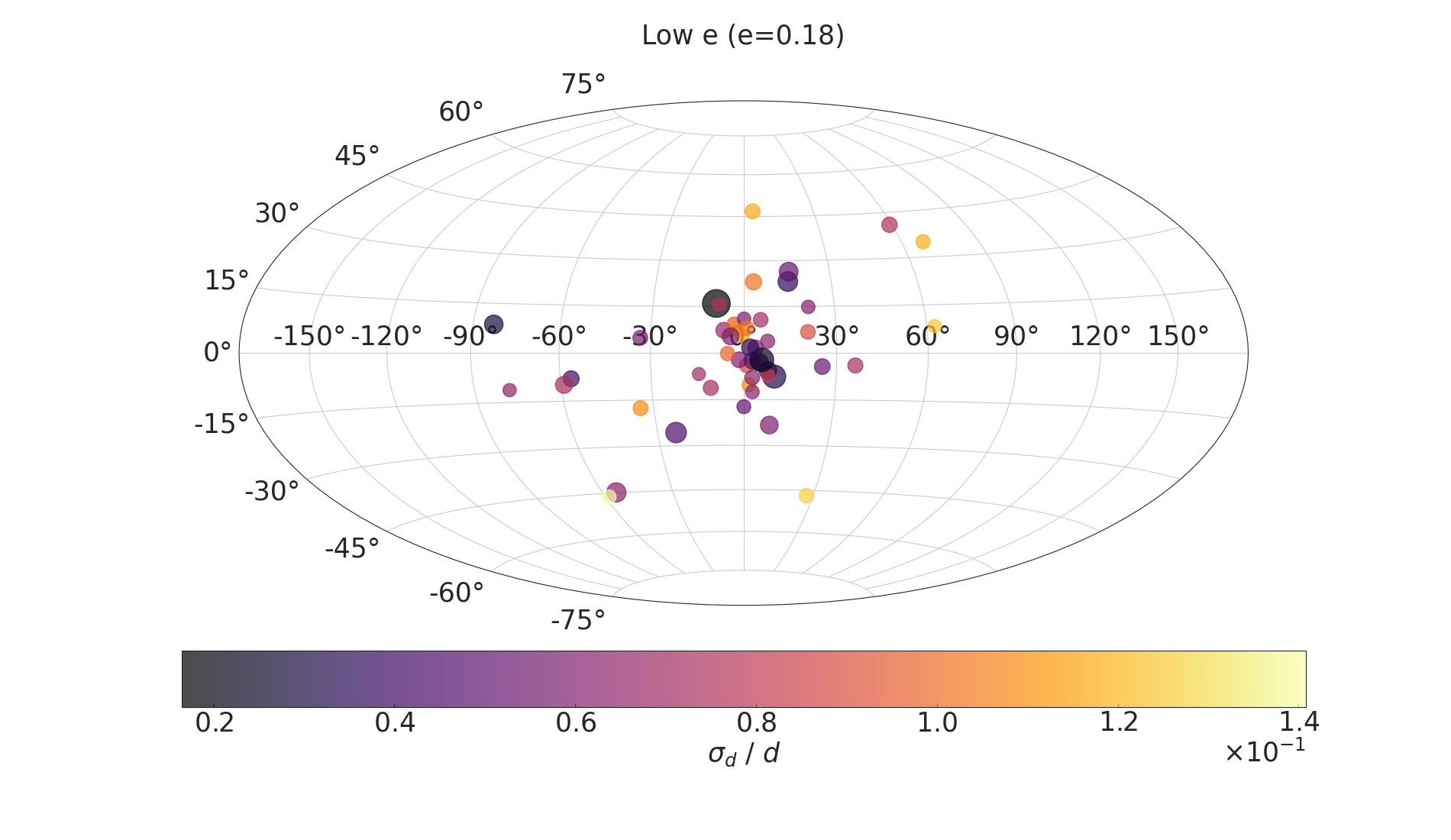}
        \label{fig:s1_dist}
    \end{subfigure}
    \hfill
    \begin{subfigure}[t]{0.48\textwidth}
        \centering
        \includegraphics[width=\linewidth]{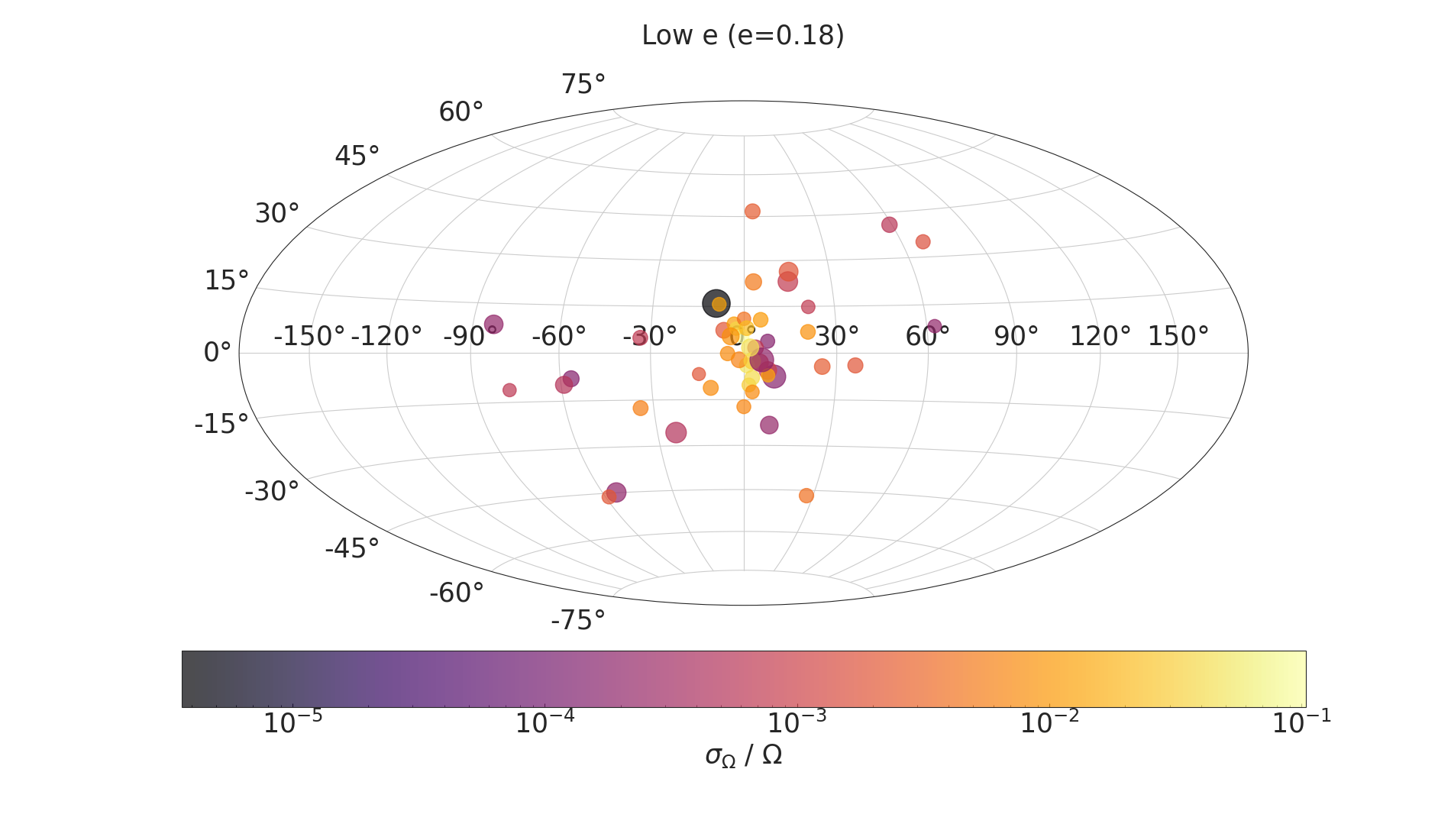}
        \label{fig:s1_loca}
    \end{subfigure}

    \vspace{1.2em}

    \begin{subfigure}[t]{0.48\textwidth}
        \centering
        \includegraphics[width=\linewidth]{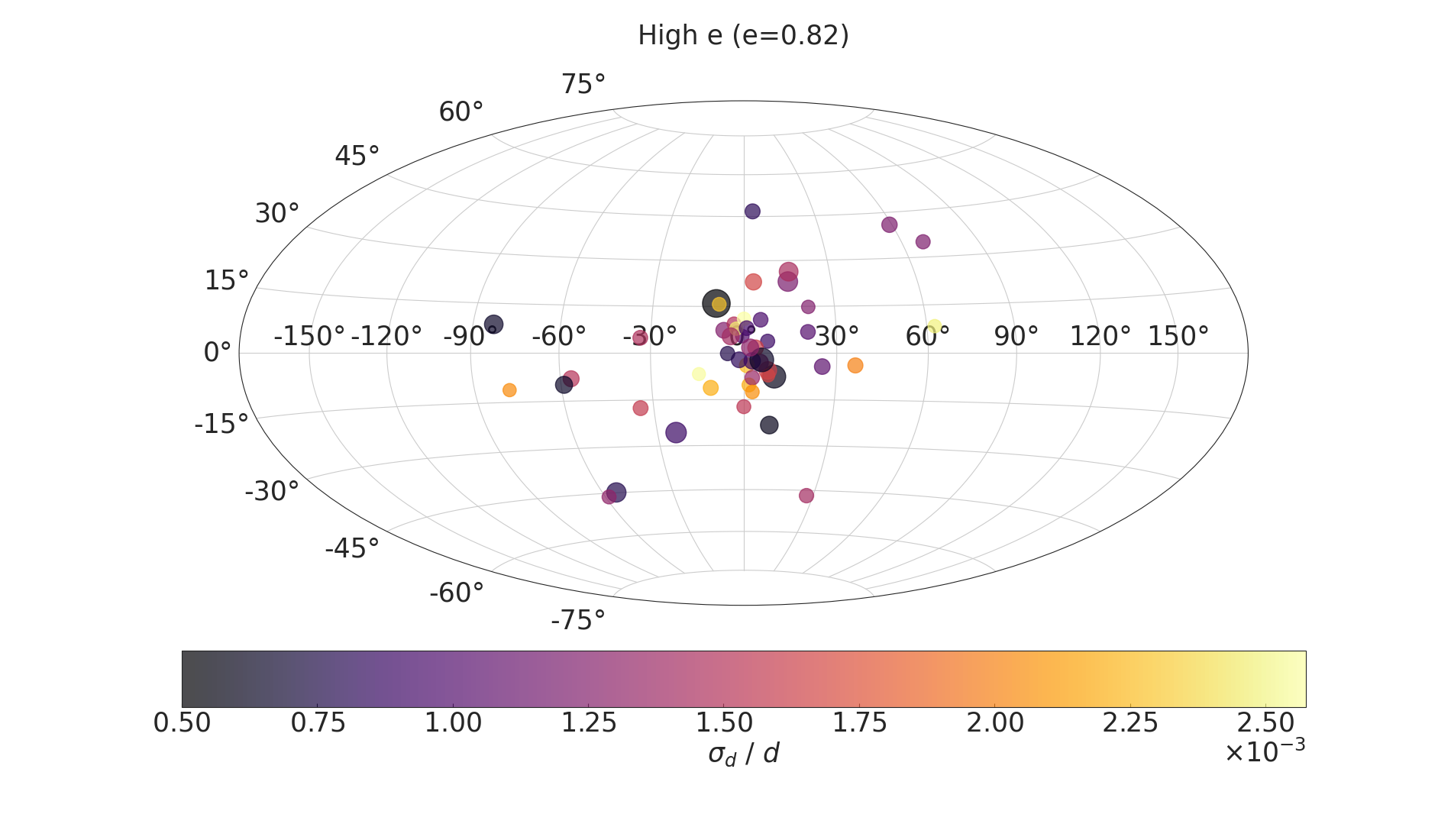}
        \label{fig:s3_dist}
    \end{subfigure}
    \hfill
    \begin{subfigure}[t]{0.48\textwidth}
        \centering
        \includegraphics[width=\linewidth]{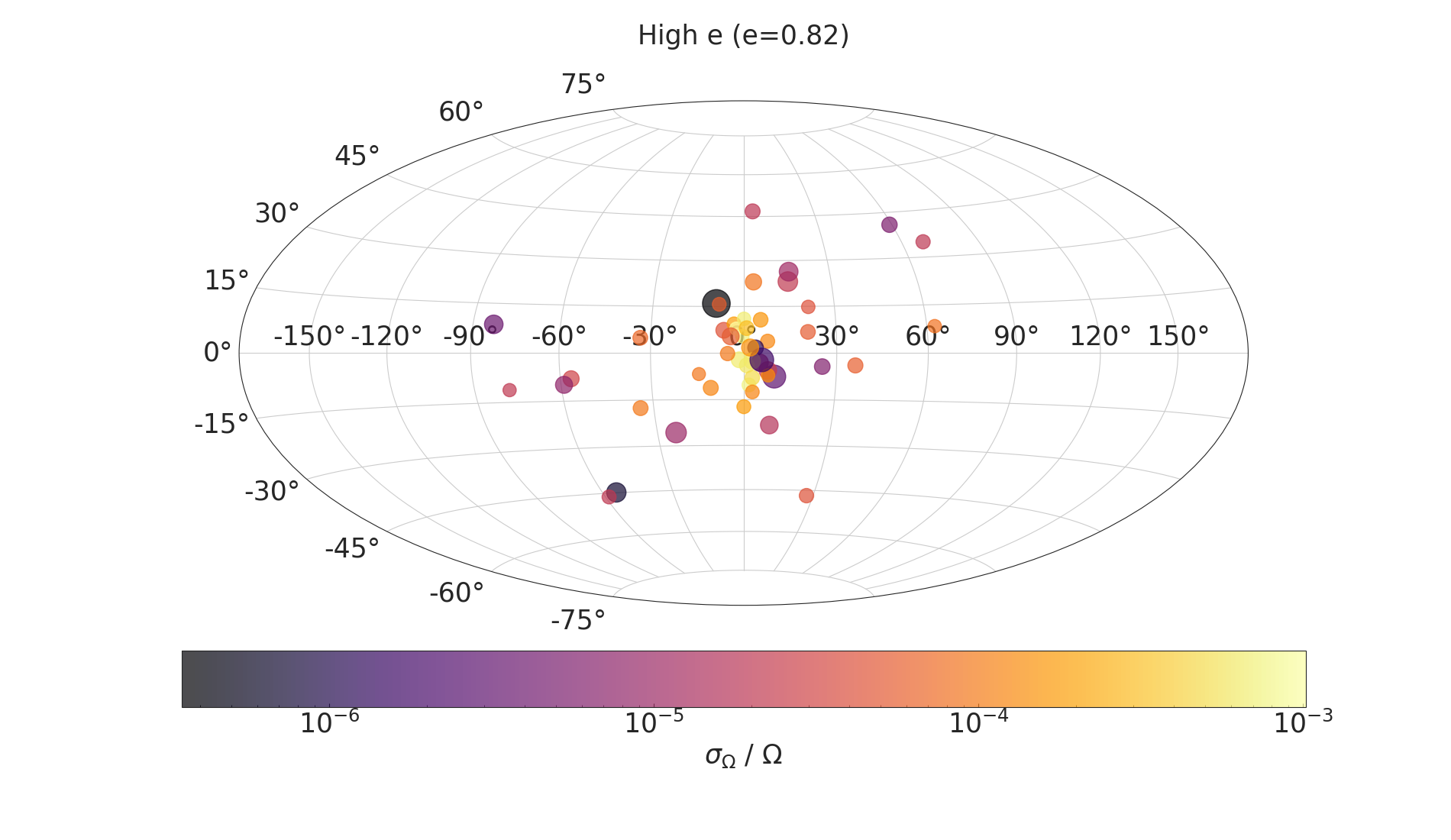}
        \label{fig:s3_loca}
    \end{subfigure}

    \vspace{1.2em}

    \begin{subfigure}[t]{0.48\textwidth}
        \centering
        \includegraphics[width=\linewidth]{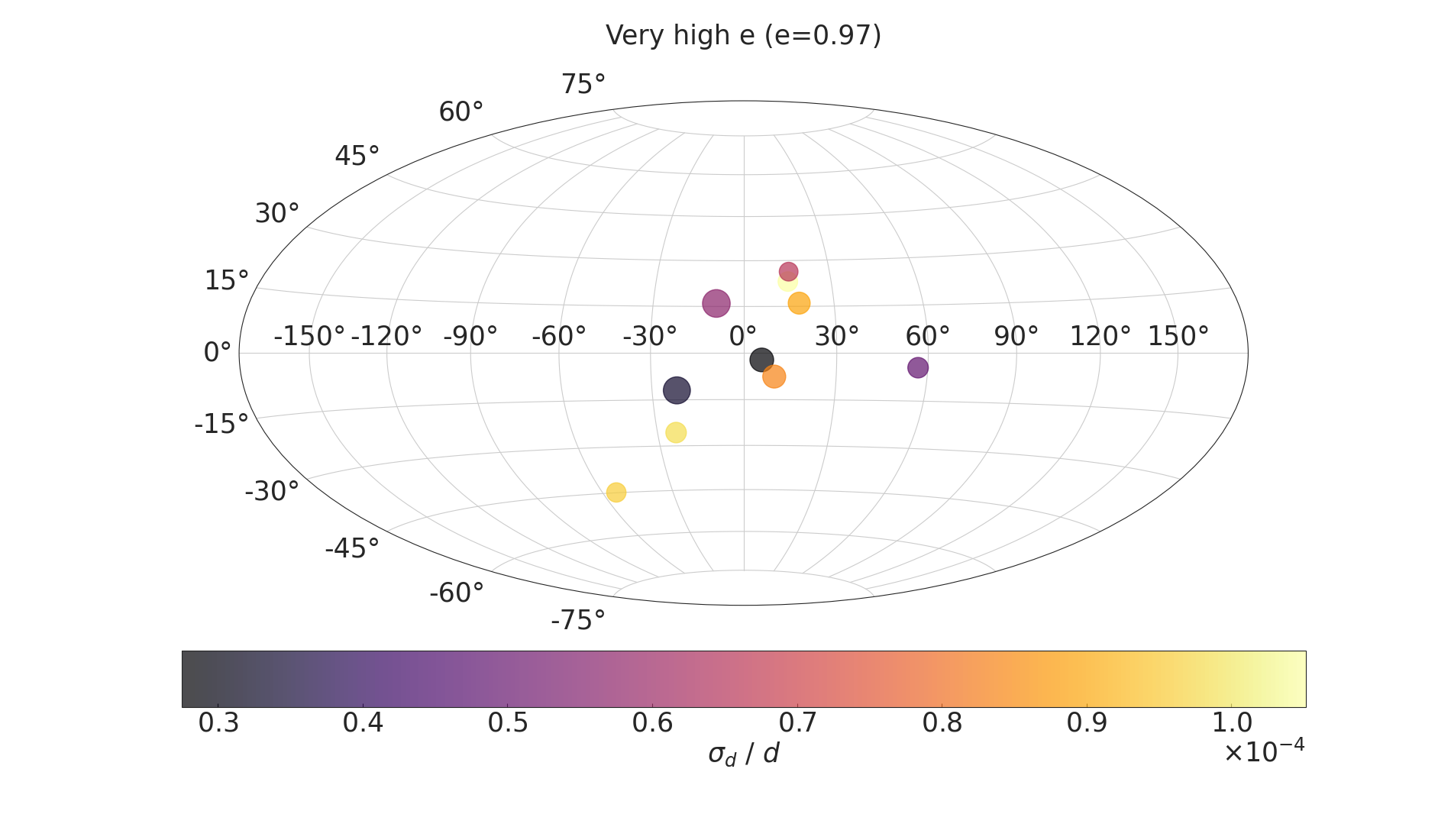}
        \label{fig:s4_dist}
    \end{subfigure}
    \hfill
    \begin{subfigure}[t]{0.48\textwidth}
        \centering
        \includegraphics[width=\linewidth]{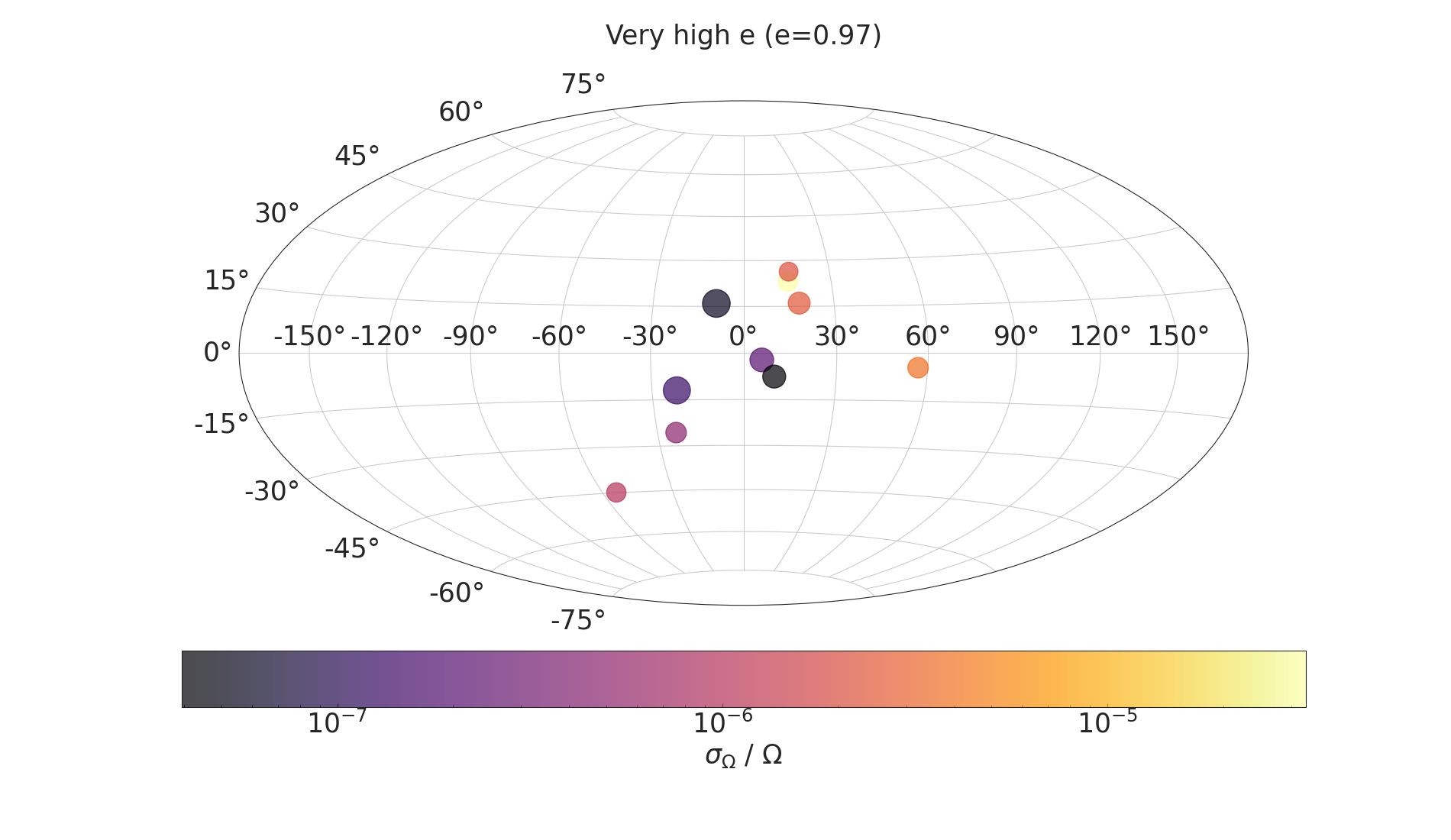}
        \label{fig:s4_loca}
    \end{subfigure}

    \vspace{1.2em}

    \begin{subfigure}[t]{0.48\textwidth}
        \centering
        \includegraphics[width=\linewidth]{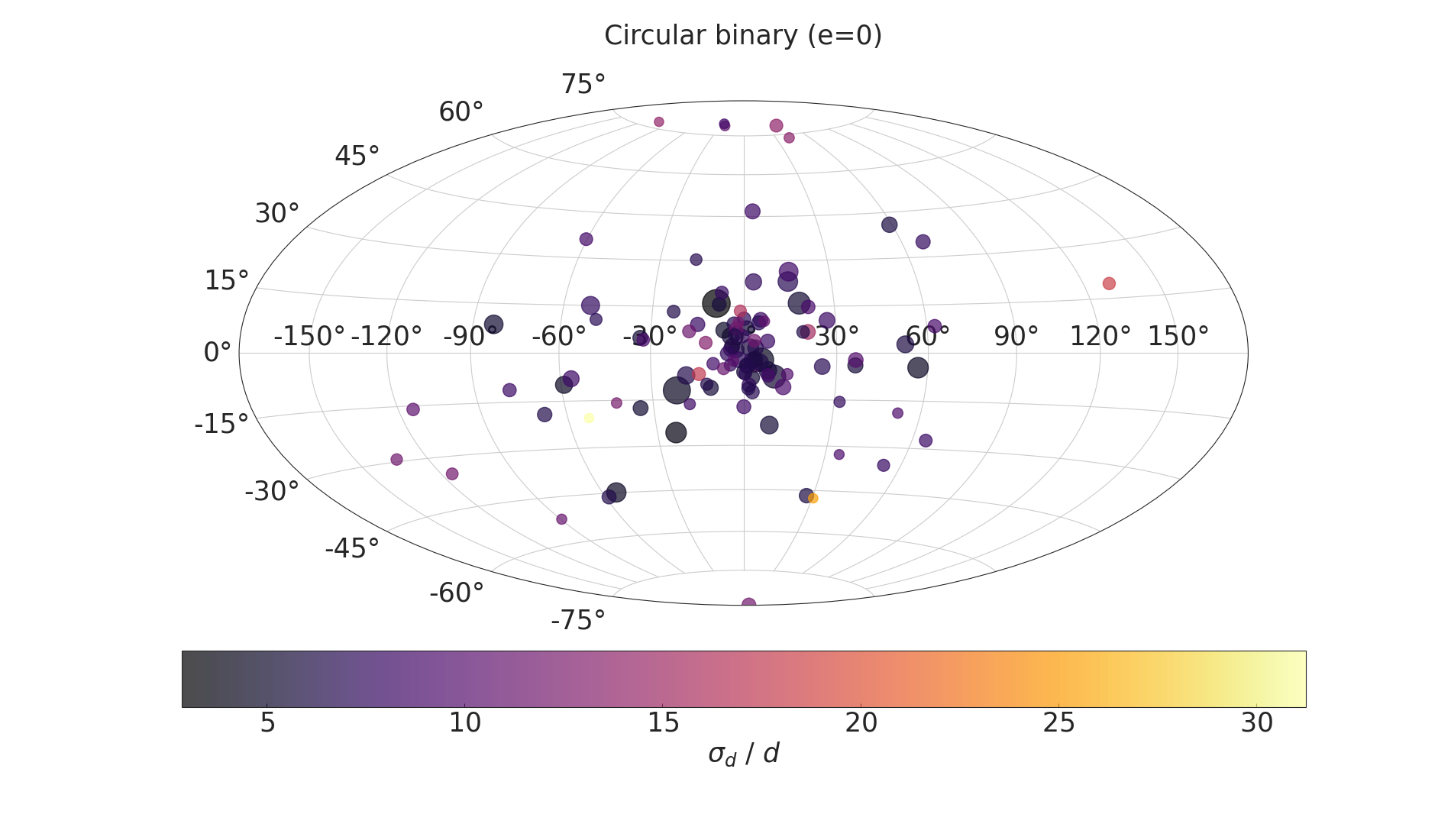}
        \label{fig:s5_dist}
    \end{subfigure}
    \hfill
    \begin{subfigure}[t]{0.48\textwidth}
        \centering
        \includegraphics[width=\linewidth]{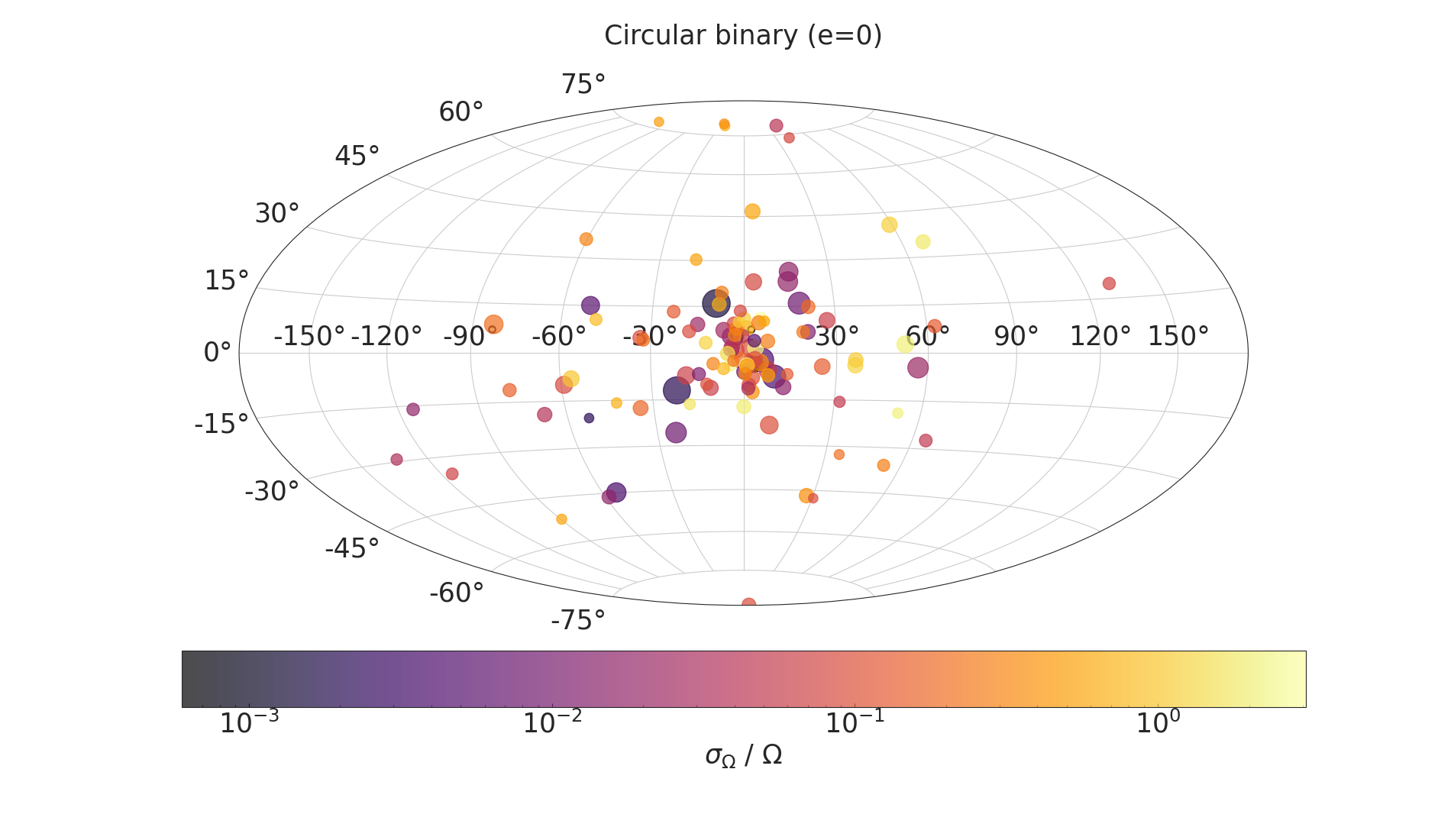}
        \label{fig:s5_loca}
    \end{subfigure}

    \caption{Skymaps of our 4 binaries placed inside different clusters. In the left column the color bar shows the distance relative error and in the right column the color bar shows the sky location relative error. Each row shows 1 binary and a larger marker indicates a smaller distance to the cluster. The sky maps are in galactic coordinates. }
    \label{fig:skymapsAccuracy}
\end{figure*}

Figure \ref{fig:skymapsAccuracy} shows skymaps of our binaries with the precision on the distance (left column) and area of the sky (localization, right column). Each row shows 1 binary in the order low e, high e, very high e, circular binary. The maps are in galactic coordinates with longitude on the x axis and inclination on the y axis. The color bar shows the precision of the selected parameter and the marker size shows the distance to the cluster where a larger marker means a smaller distance. A look at both columns shows that many GCs are positioned around the galactic bulge as expected with a few clusters spread around in the halo.

Focusing on the precision of the distance estimation (left column) we can see that, as already shown, the precision gets better with higher eccentricity with quite poor precision for circular binaries. We do not see any dependence on position in the MW for the precision on the distance estimation with both large and small precisions at different locations. 

The precision on the location estimate follows the same pattern of higher eccentricity gives better precision. There is a larger spread between the different clusters due to the difference in area on the sky of each cluster. More distant clusters have a much smaller sky area than closer clusters which gives a large range in the relative precision while the absolute precisions are fairly similar. With a high eccentricity the precision is high enough to be able to determine even the position of the binary inside the cluster, this is only possible for the very closest clusters for the circular or low eccentricity binary.

\subsection{GW and EM signals}
\label{sec:mtdwd}

Although not central to our main analysis, we include this exploratory study
to assess the potential for complementary EM identification of GW sources.

Mass-transferring DWDs are potential EM sources, since accretion occurs when
the binary components come into close proximity
\citep{Rueda2019, Rueda2013}. However, many of these sources are intrinsically
faint, making them difficult to identify \citep[e.g.][]{Ramsay2018} and
precisely localize using EM observations alone. GWs provide an independent
probe unaffected by interstellar gas or dust and offer a complementary means
of studying these systems
\citep{Breivik2018, korol2022, Jin2025, korol2017}. Combining GW and EM
observations is therefore expected to yield substantially improved constraints
on mass-transferring DWDs.

In this work, we conduct a preliminary test to assess whether joint EM and GW observations is possible for mass transferring DWDs. We use the MOCCA code described in Section~\ref{sec:models} \citep{Hypki2013, Giersz2013, hypki2025, giersz2025} to simulate mass transferring DWDs populations of different ages and apply the FIM to estimate GW-based localization uncertainties \citep{Cutler1994, Grimm2020, Liu2020}. For demonstration purpose, we assume sources are in a globular cluster with distance 2 \rm{kpc} from us and the tidal radius of the cluster is 50 \rm{pc} which is a common scales for globular clusters \citep{van2008, Portegies2010}. In our simulation, we analyze several evolutionary snapshots of the globular cluster simply to demonstrate the localization ability for  mass-transferring DWDs across different late evolutionary stages. 
We first select binaries older than 9 \rm{Gyrs}, and then divide them into 250 \rm{Myrs}-wide snapshots centered at 9, 10, 11 and 12 \rm{Gyrs}. 
For sources older than 9 \rm{Gyrs}, $180/30$ systems achieve sky areas $\Delta \Omega\leq10/1\,\rm{deg^2}$. For 9-Gyr systems, $12/0$ reach $\Delta \Omega\leq10/1 \rm{deg^2}$; for 10-Gyr systems, $7/0$; for 11-Gyr systems, $18/4$; and for 12-Gyr systems, $20/8$. Compared with the fields of view of current EM observatories, such as Gaia ($1.26 \rm{deg^2}$) \citep{Gaia2016, Gaia_D3} and VRO ($9.6 \rm{deg^2}$) \citep{LSST}, a considerable number of sources are expected to be detectable by both GW and EM, indicating that joint GW+EM observations are possible. As a preliminary study, we only explore the feasibility of combining GW and EM observations. In future work, we will show the detailed method for joint EM-GW analysis and show how it can used to reveal the properties of AM CVn binaries. 

\section{Conclusions and discussion}
\label{sec:discussion}

We study the parameter estimation uncertainties of DWDs in globular clusters that may be detected by LISA in the future. We consider both eccentric and circular DWDs from MOCCA models and use sky position and distances to real clusters in the MW. We find that for highly eccentric binaries, the relative sky localization (normalized by their host cluster area) could be about $10^{-7}-0.001$, even for binaries with low eccentricity ($e < 0.2$) the relative sky localization can be smaller than $0.1$. Indicating that we could localize the eccentric DWDs inside the cluster, and even for circular binaries it can be localized inside their host cluster for plenty of cluster positions. For the luminosity distance, binaries with high eccentricity can achieve the relative precision of $10^{-5}-0.003$, while for the binaries with small eccentricity and circular binaries the relative error of luminosity distance is in the range of $0.01-30$. 
Compared to current distance-estimation methods, we find that binaries with significant eccentricity yield precisions several orders of magnitude higher.
Current distance-estimation methods typically achieve, at best, $\sim 1\%$ precision \citep{baumgardt2021}, while our results suggest potential improvements of roughly 2–3 orders of magnitude in the most favorable cases. Even a DWD with low eccentricity shows precision similar to current day methods at closer distances.  Tight and eccentric DWDs can be used to obtain very accurate estimations on both distance and sky location, providing a very valuable and independent tool for the distance ladder.
For circular DWDs inside GCs, their sky localizations are not particularly accurate compared to eccentric DWDs and existing methods, however, due to the large number of them they might still provide useful estimations. We find that a number of mass-transferring DWDs in globular clusters are expected to be detectable in both GW and EM signals, offering a new opportunity to probe the detailed properties of these systems.

Note that we adopt the FIM method when estimating the parameter uncertainties through the paper rather than a full Bayesian approach \citep{2014PhRvD..89d2004G}. Since FIM is only valid under the situation of high SNR limit when the non-linear terms of the signals can be ignored \citep{2008PhRvD..77d2001V}, and it do not contain the systematical errors from the waveform template. Thus our estimate gives an optimistic uncertainties for the DWDs in clusters. For example, the estimations on frequencies depend on the observation period and the cadence of data in the realistic cases, for highly eccentric sources in a more realistic situation, the relative errors for the frequency may be much larger than the results presented here \citep{2025ApJ...981...66D}. Roughly speaking, for localization FIM method could also produce an order of magnitude overestimation \citep{2013PhRvD..88h4013R}.

Several simplifying assumptions were adopted in this work, including perfect knowledge of cluster positions and the use of representative simulated binaries rather than a full population synthesis. These choices may affect quantitative precision estimates, but they are not expected to alter the qualitative trends identified here or the applicability of the proposed distance-estimation methodology.

A challenge for this method is to isolate the signal from the eccentric DWD from a large number of other signals in the same frequency range. \cite{relton2022} bring up techniques that can be used with current day detectors (LIGO, Virgo, KAGRA) to isolate signals from overlapping binary black hole mergers without significant losses to the signal. For future detectors and weaker signals (such as those from DWDs) there has been research done to find ways of isolating the signals \citep[e.g.]{niu2024, houba2025, Littenberg2020} and, given recent advances in signal-analysis techniques, this is unlikely to represent a fundamental limitation for the proposed distance-estimation method, although a detailed assessment is beyond the scope of this work. These binaries are also emitting GWs stronger than most binaries at this frequency.

Another question is whether there is enough eccentric and tight DWDs in the MW at a given time. This type of binary is rare and requires a specific formation channel to form. Nevertheless an expected upper limit is between 10-15 in the MW \citep{hellstrom2025} at any given time. Even though this is a low number, the high precision on the parameter estimation would give important constrains for the distance scale and can provide significant insight into the distances to GCs in the MW. GW distance measurements are unaffected by dust extinction and can therefore be used to cross-calibrate traditional standard candles, constrain brightness-metallicity relations for RR Lyrae stars or the TRGB, and improve reddening estimates, which affect nearly all standard candles.

\begin{figure*}
    \centering

    \begin{subfigure}[t]{0.48\textwidth}
        \centering
        \includegraphics[width=\linewidth]{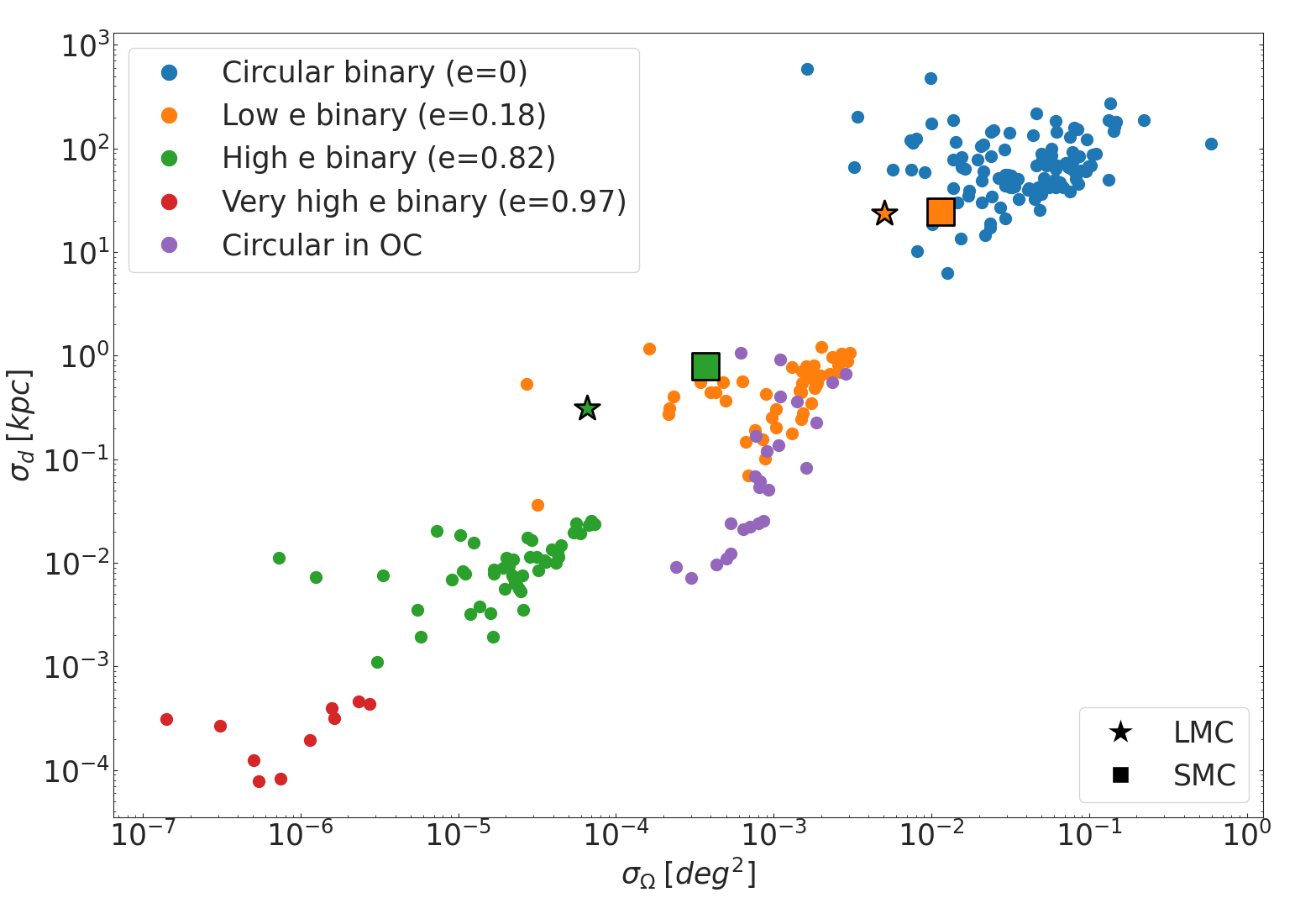}
        \caption{}
        \label{fig:OCScatterPercent}
    \end{subfigure}
    \hfill
    \begin{subfigure}[t]{0.48\textwidth}
        \centering
        \includegraphics[width=\linewidth]{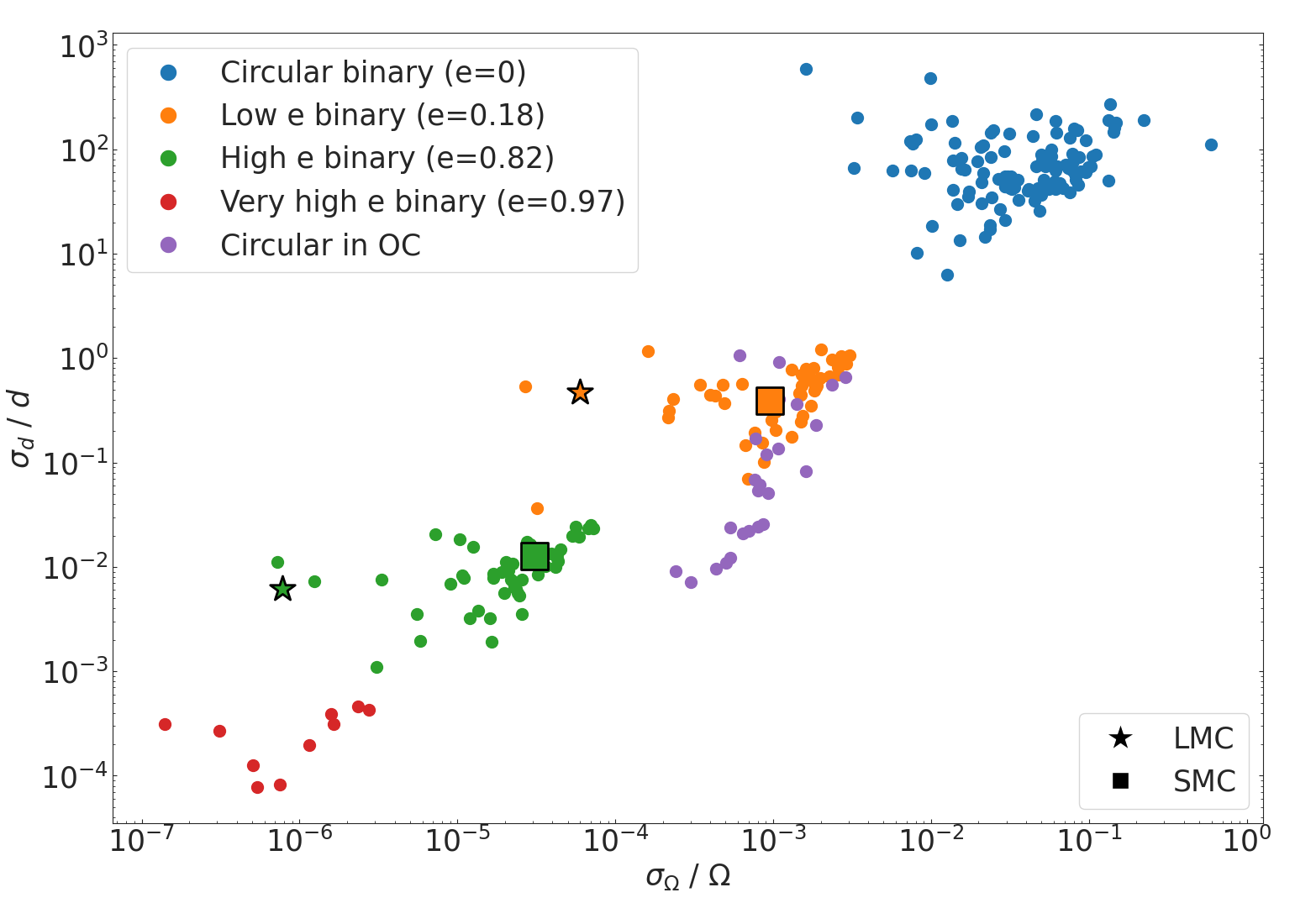}
        \caption{}
        \label{fig:OCScatterAbsolute}
    \end{subfigure}

    \caption{Same scatter plots as in figure \ref{fig:scatterAcc} but with the 5 binaries placed in OCs added as purple circles and two binaries in LMC (stars) and SMC (squares) respectively.}
    \label{fig:scatterOC}
\end{figure*}

As discussed in section \ref{sec:skyLocaAndDistResults} the precision from circular binaries inside GCs are generally not good, however, circular binaries can also be found in the field and in open clusters. We took 5 additional tight circular binaries and put them at the Hyades, Coma Berenices, Pleiades, Praesepe and Ruprecht 147 clusters. The precisions can be seen in figure \ref{fig:scatterOC} where we show the precisions for the 4 groups as previously shown in the section \ref{sec:skyLocaAndDistResults} together with the 5 circular binaries placed in the open clusters (purple circles). In the first panel  we show the absolute error and we can see that the binaries in the OCs have similar precision in the location estimate to the low $e$ binaries. However, the distance estimate is still worse than all eccentric binaries, although better than the circular binaries inside GCs as expected. Looking at the relative precision (second panel) we can see that although the absolute error in distance is low for the binaries in OCs, the relative precision is worse than all other groups due to the small distance to the source and thus a large area of the sky of the OC. Due to the large error on distance, it would be difficult to say whether the binary is inside the OC rather than behind or in front, however \cite{vanZeist2025} shows that for a location precision this high, we should not expect any field binaries to be in the same sky location. 

We also investigate whether this method could be used to investigate the distance to the LMC and the SMC. The low eccentricity and the high eccentricity binary is placed in our cloud and their signal is simulated. We use the area of the whole cloud, not just a GC inside the cloud. The distance and location estimations can be seen in figure \ref{fig:scatterOC} as stars (LMC) and squares (SMC). From this we can see that the precision is good, however, not as good as inside the MW GCs. Nevertheless, from our results, assuming that there is at least one tight and eccentric DWD inside either of the Magellanic clouds we should be able to use the GW signal from such a binary to constrain our distance measurements to them.

\begin{acknowledgements}
LH, MG, AA, GW were supported by the Polish National Science Center (NCN) through the grant 2021/41/B/ST9/01191.
AA acknowledges that this research was funded in part by National Science Centre (NCN), Poland, grant No. 2024/55/D/ST9/02585. For the purpose of Open Access, the author has applied a CC BY public copyright licence to any Author Accepted Manuscript (AAM) version arising from this submission. 
DB acknowledges support from the São Paulo Research Foundation (FAPESP), Brazil, Process Numbers 2024/03736-2 and 2025/00817-4.
YL is partly supported by the National Key Program for Science and Technology Research and Development (grant no. 2020YFC2201400 and 2022YFC2205201) and the National Astronomical Observatory of China (grant No. E4TG660101).
YZ acknowledges support from the Postdoctoral Fellowship Program of CPSF (GZC20240127).
\end{acknowledgements}

\section*{Data Availability}
Input and output data for the globular cluster simulations carried out in this paper is available to download\footnote{\url{https://zenodo.org/records/10865904}}.

\bibliographystyle{aa}
\bibliography{ref.bib}

\clearpage
\onecolumn
\section*{Appendix A}
\label{sec:appendixA}
    \begin{longtable}{c c c c c c}
        \caption{Table of GC parameters obtained from \citet{2018MNRAS.478.1520B} that were used for all binaries. The cluster name in bold shows that the cluster were used for eccentric binaries, among these clusters a star next to the name shows the 10 binaries that we used for the very high eccentricity binary (e=0.97).}      
        \label{tab:gcs2} \\  
        
        \hline\hline
            Name & Mass [M$_{\odot}$] & Distance [kpc] & RA [deg] & Dec [deg] & area [sq deg]\\
        \hline
        \endfirsthead
        \caption{continued.}\\
        \hline\hline
            Name & Mass [M$_{\odot}$] & Distance [kpc] & RA [deg] & Dec [deg] & area [sq deg]\\
            \hline
        \endhead
        \hline
        \endfoot
            \textbf{NGC6121*} & 1.29e+05 & 1.85 & 245.9 & -26.53 & 8.77 \\ 
            NGC6397 & 7.75e+04 & 2.48 & 265.18 & -53.67 & 4.13 \\ 
            \textbf{NGC6544*} & 1.02e+05 & 2.58 & 271.84 & -25.0 & 3.53 \\ 
            \textbf{NGC6656} & 4.30e+05 & 3.3 & 279.1 & -23.9 & 5.59 \\ 
            NGC6366 & 3.38e+04 & 3.44 & 261.93 & -5.08 & 1.25 \\ 
            NGC6838 & 3.00e+04 & 4.0 & 298.44 & 18.78 & 1.08 \\ 
            \textbf{NGC6752*} & 2.11e+05 & 4.12 & 287.72 & -59.98 & 2.68 \\ 
            \textbf{NGC104*} & 1.00e+06 & 4.52 & 6.02 & -72.08 & 7.84 \\ 
            \textbf{NGC3201} & 1.63e+05 & 4.74 & 154.4 & -46.41 & 3.11 \\ 
            \textbf{NGC6254} & 1.68e+05 & 5.07 & 254.29 & -4.1 & 1.16 \\ 
            \textbf{NGC6218} & 1.44e+05 & 5.11 & 251.81 & -1.95 & 0.82 \\ 
            \textbf{NGC6553} & 2.19e+05 & 5.33 & 272.32 & -25.91 & 0.83 \\ 
            \textbf{NGC6809} & 1.82e+05 & 5.35 & 295.0 & -30.96 & 0.98 \\ 
            \textbf{NGC6626} & 3.14e+05 & 5.37 & 276.14 & -24.87 & 0.8 \\ 
            NGC5139 & 3.6e+06 & 5.43 & 201.7 & -47.48 & 15.22 \\ 
            NGC6352 & 6.62e+04 & 5.54 & 261.37 & -48.42 & 0.33 \\ 
            \textbf{NGC6171} & 1.21e+05 & 5.63 & 248.13 & -13.05 & 0.4 \\ 
            Terzan1 & 9.93e+03 & 5.67 & 263.95 & -30.47 & 0.47 \\ 
            \textbf{NGC4372*} & 2.23e+05 & 5.71 & 186.44 & -72.66 & 1.73 \\ 
            Terzan9 & 5.21e+03 & 5.77 & 270.41 & -26.84 & 0.31 \\ 
            NGC6540 & 5.93e+04 & 5.91 & 271.54 & -27.77 & 0.15 \\ 
            \textbf{NGC6266} & 8.04e+05 & 6.03 & 255.3 & -30.11 & 0.75 \\ 
            \textbf{NGC6304} & 1.42e+05 & 6.15 & 258.63 & -29.46 & 0.2 \\ 
            NGC6535 & 1.36e+04 & 6.36 & 270.96 & -0.3 & 0.16 \\ 
            \textbf{NGC4833} & 3.17e+05 & 6.48 & 194.89 & -70.88 & 1.46 \\ 
            \textbf{Terzan5} & 1.59e+05 & 6.62 & 267.02 & -24.78 & 0.62 \\ 
            \textbf{Pal6} & 8.89e+04 & 7.05 & 265.93 & -26.22 & 0.02 \\ 
            NGC6256 & 1.24e+05 & 7.24 & 254.89 & -37.12 & 0.15 \\ 
            \textbf{Terzan6} & 1.86e+05 & 7.27 & 267.7 & -31.27 & 0.04 \\ 
            \textbf{NGC6522} & 1.96e+05 & 7.29 & 270.89 & -30.03 & 0.05 \\ 
            \textbf{NGC6712} & 1.71e+05 & 7.38 & 283.27 & -8.71 & 0.21 \\ 
            \textbf{NGC6205*} & 4.50e+05 & 7.42 & 250.42 & 36.46 & 3.18 \\ 
            NGC6401 & 2.47e+05 & 7.44 & 264.65 & -23.91 & 0.05 \\ 
            \textbf{NGC5904} & 5.72e+05 & 7.48 & 229.64 & 2.08 & 1.22 \\ 
            NGC6717 & 3.14e+04 & 7.52 & 283.78 & -22.7 & 0.06 \\ 
            \textbf{NGC6325} & 1.04e+05 & 7.53 & 259.5 & -23.77 & 0.08 \\ 
            Terzan4 & 1.06e+04 & 7.59 & 262.66 & -31.6 & 0.04 \\ 
            NGC6749 & 8.19e+04 & 7.59 & 286.31 & 1.9 & 0.71 \\ 
            \textbf{NGC6541*} & 4.38e+05 & 7.61 & 272.01 & -43.71 & 0.25 \\ 
            Terzan3 & 1.45e+04 & 7.64 & 247.17 & -35.35 & 0.08 \\ 
            \textbf{NGC6362} & 1.03e+05 & 7.65 & 262.98 & -67.05 & 0.45 \\ 
            Terzan2 & 3.85e+04 & 7.75 & 261.89 & -30.8 & 0.03 \\ 
            NGC6558 & 6.44e+04 & 7.79 & 272.57 & -31.76 & 0.01 \\ 
            NGC6528 & 7.26e+04 & 7.83 & 271.21 & -30.06 & 0.02 \\ 
            E3 & 7.60e+03 & 7.88 & 140.24 & -77.28 & 0.07 \\ 
            \textbf{NGC6287} & 1.50e+05 & 7.93 & 256.29 & -22.71 & 0.08 \\ 
            NGC6342 & 6.32e+04 & 8.01 & 260.29 & -19.59 & 0.04 \\ 
            \textbf{NGC6624} & 1.69e+05 & 8.02 & 275.92 & -30.36 & 0.05 \\ 
            NGC6642 & 7.89e+04 & 8.05 & 277.98 & -23.48 & 0.04 \\ 
            \textbf{Liller1} & 1.45e+05 & 8.06 & 263.35 & -33.39 & 0.24 \\ 
            \textbf{NGC6144} & 9.40e+04 & 8.15 & 246.81 & -26.02 & 0.13 \\ 
            \textbf{NGC6539} & 3.54e+05 & 8.16 & 271.21 & -7.59 & 0.31 \\ 
            \textbf{NGC6440} & 5.41e+05 & 8.25 & 267.22 & -20.36 & 0.17 \\ 
            \textbf{NGC5927} & 2.28e+05 & 8.27 & 232.0 & -50.67 & 0.63 \\ 
            \textbf{NGC6723} & 2.32e+05 & 8.27 & 284.89 & -36.63 & 0.22 \\ 
            \textbf{NGC6333} & 2.59e+05 & 8.3 & 259.8 & -18.52 & 0.17 \\ 
            \textbf{NGC6273} & 7.67e+05 & 8.34 & 255.66 & -26.27 & 0.23 \\ 
            \textbf{NGC6760} & 2.34e+05 & 8.41 & 287.8 & 1.03 & 0.79 \\ 
            \textbf{NGC7099} & 1.63e+05 & 8.46 & 325.09 & -23.18 & 0.73 \\ 
            \textbf{NGC6341} & 3.29e+05 & 8.5 & 259.28 & 43.14 & 1.82 \\ 
            \textbf{NGC6355} & 2.89e+05 & 8.65 & 260.99 & -26.35 & 0.03 \\ 
            \textbf{Djorg2} & 1.08e+05 & 8.76 & 270.45 & -27.83 & 0.03 \\ 
            \textbf{NGC362} & 4.03e+05 & 8.83 & 15.81 & -70.85 & 1.0 \\ 
            \textbf{NGC6637} & 1.95e+05 & 8.9 & 277.85 & -32.35 & 0.08 \\ 
            Pal10 & 3.54e+04 & 8.94 & 289.51 & 18.57 & 0.52 \\ 
            NGC288 & 8.57e+04 & 8.99 & 13.19 & -26.58 & 1.16 \\ 
            \textbf{NGC6402} & 7.47e+05 & 9.14 & 264.4 & -3.25 & 0.68 \\ 
            \textbf{NGC6293*} & 2.21e+05 & 9.19 & 257.54 & -26.58 & 0.07 \\ 
            NGC6517 & 3.41e+05 & 9.23 & 270.46 & -8.96 & 0.27 \\ 
            \textbf{NGC6681} & 1.21e+05 & 9.36 & 280.8 & -32.29 & 0.09 \\ 
            NGC6652 & 7.89e+04 & 9.46 & 278.94 & -32.99 & 0.04 \\ 
            NGC6380 & 1.71e+05 & 9.61 & 263.62 & -39.07 & 0.18 \\ 
            NGC6496 & 1.30e+05 & 9.64 & 269.77 & -44.27 & 0.09 \\ 
            NGC5946 & 1.27e+05 & 9.64 & 233.87 & -50.66 & 0.25 \\ 
            \textbf{NGC6638} & 1.21e+05 & 9.78 & 277.73 & -25.5 & 0.09 \\ 
            NGC2298 & 5.72e+04 & 9.83 & 102.25 & -36.01 & 0.57 \\ 
            Djorg1 & 1.06e+05 & 9.88 & 266.87 & -33.07 & 0.07 \\ 
            NGC6139 & 3.78e+05 & 10.04 & 246.92 & -38.85 & 0.35 \\ 
            \textbf{NGC2808*} & 9.75e+05 & 10.06 & 138.01 & -64.86 & 2.63 \\ 
            NGC6453 & 1.32e+05 & 10.07 & 267.72 & -34.6 & 0.11 \\ 
            NGC5272 & 6.10e+05 & 10.18 & 205.55 & 28.38 & 1.66 \\ 
            NGC6093 & 3.35e+05 & 10.34 & 244.26 & -22.98 & 0.38 \\ 
            NGC4590 & 1.52e+05 & 10.4 & 189.87 & -26.74 & 0.58 \\ 
            \textbf{NGC6779} & 1.57e+05 & 10.43 & 289.15 & 30.18 & 0.81 \\ 
            NGC6569 & 3.51e+05 & 10.53 & 273.41 & -31.83 & 0.17 \\ 
            NGC5986 & 4.06e+05 & 10.54 & 236.51 & -37.79 & 0.44 \\ 
            NGC7078 & 8.11e+05 & 10.71 & 322.49 & 12.17 & 1.57 \\ 
            NGC5286 & 5.36e+05 & 11.1 & 206.61 & -51.37 & 0.76 \\ 
            NGC6316 & 3.71e+05 & 11.15 & 259.16 & -28.14 & 0.19 \\ 
            \textbf{NGC6388*} & 9.93e+05 & 11.17 & 264.07 & -44.74 & 0.78 \\ 
            Pal1 & 1.74e+03 & 11.18 & 53.33 & 79.58 & 0.04 \\ 
            Pal8 & 2.74e+04 & 11.32 & 280.37 & -19.83 & 0.11 \\ 
            NGC7089 & 7.00e+05 & 11.69 & 323.36 & -0.82 & 0.93 \\ 
            NGC6235 & 5.61e+04 & 11.94 & 253.36 & -22.18 & 0.1 \\ 
            NGC1851 & 3.67e+05 & 11.95 & 78.53 & -40.05 & 1.1 \\ 
            NGC5897 & 1.33e+05 & 12.55 & 229.35 & -21.01 & 0.34 \\ 
            NGC6441 & 1.22e+06 & 12.73 & 267.55 & -37.05 & 0.75 \\ 
            NGC1904 & 2.38e+05 & 13.08 & 81.04 & -24.52 & 0.26 \\ 
            NGC6584 & 2.04e+05 & 13.61 & 274.66 & -52.22 & 0.2 \\ 
            Pal11 & 1.00e+05 & 14.02 & 296.31 & -8.01 & 0.06 \\ 
            NGC6284 & 2.61e+05 & 14.21 & 256.12 & -24.76 & 0.27 \\ 
            NGC6101 & 1.02e+05 & 14.45 & 246.45 & -72.2 & 0.39 \\ 
            NGC6356 & 4.34e+05 & 15.66 & 260.9 & -17.81 & 0.59 \\ 
            NGC6934 & 1.63e+05 & 15.72 & 308.55 & 7.4 & 0.33 \\ 
            NGC5466 & 1.06e+05 & 16.12 & 211.36 & 28.53 & 0.23 \\ 
            NGC1261 & 2.25e+05 & 16.4 & 48.07 & -55.22 & 0.73 \\ 
            NGC6981 & 1.12e+05 & 16.66 & 313.37 & -12.54 & 0.11 \\ 
            NGC5053 & 8.65e+04 & 17.54 & 199.11 & 17.7 & 0.28 \\ 
            Pal12 & 1.05e+04 & 18.49 & 326.66 & -21.25 & 0.12 \\ 
            NGC5024 & 5.21e+05 & 18.5 & 198.23 & 18.17 & 1.01 \\ 
            NGC4147 & 5.02e+04 & 18.54 & 182.53 & 18.54 & 0.26 \\ 
            IC4499 & 1.45e+05 & 18.89 & 225.08 & -82.21 & 0.35 \\ 
            NGC6864 & 4.58e+05 & 20.52 & 301.52 & -21.92 & 0.6 \\ 
            NGC6426 & 7.96e+04 & 20.71 & 266.23 & 3.17 & 0.19 \\ 
            Rup106 & 5.93e+04 & 20.71 & 189.67 & -51.15 & 0.13 \\ 
            Pal5 & 2.00e+04 & 21.94 & 229.02 & -0.11 & 0.05 \\ 
            Pal13 & 5.46e+03 & 23.48 & 346.69 & 12.77 & 0.03 \\ 
            Terzan7 & 1.73e+04 & 24.28 & 289.43 & -34.66 & 0.07 \\ 
            NGC7492 & 3.61e+04 & 24.39 & 347.11 & -15.61 & 0.13 \\ 
            NGC5634 & 2.04e+05 & 25.96 & 217.41 & -5.98 & 0.57 \\ 
            Pal2 & 2.64e+05 & 26.17 & 71.52 & 31.38 & 0.22 \\ 
            NGC6715 & 1.68e+06 & 26.28 & 283.76 & -30.48 & 1.16 \\ 
            IC1257 & 4.93e+04 & 26.59 & 261.77 & -7.08 & 0.05 \\ 
            Terzan8 & 1.82e+04 & 27.54 & 295.44 & -34.0 & 0.16 \\ 
            Arp2 & 2.23e+04 & 28.73 & 292.18 & -30.36 & 0.09 \\ 
            NGC6229 & 2.86e+05 & 30.11 & 251.74 & 47.53 & 0.42 \\ 
            NGC5824 & 5.93e+05 & 31.71 & 225.99 & -33.07 & 0.67 \\ 
            NGC5694 & 2.32e+05 & 34.84 & 219.9 & -26.54 & 0.44 \\ 
            NGC7006 & 2.00e+05 & 39.32 & 315.37 & 16.19 & 0.19 \\ 
            Pal15 & 2.74e+04 & 44.1 & 254.96 & -0.53 & 0.06 \\ 
            Pal14 & 1.42e+04 & 73.58 & 242.75 & 14.96 & 0.0 \\ 
            NGC2419 & 1.00e+06 & 88.47 & 114.54 & 38.88 & 0.71 \\ 
            Pal3 & 3.23e+04 & 94.84 & 151.38 & 0.07 & 0.06 \\ \hline
    \end{longtable}
\end{document}